\documentclass[prl,aps,amsfonts,amsmath,amssymb,nofootinbib,twocolumn,superscriptaddress]{revtex4}
\usepackage{mathtools}
\usepackage{upgreek}
\usepackage{graphicx}
\usepackage{bm}
\usepackage{hyperref}
\usepackage{braket}
\newcommand{\RNum}[1]{\uppercase\expandafter{\romannumeral #1\relax}}

\usepackage{natbib}
\usepackage{float}
\usepackage{xcolor}

\restylefloat{table}

\begin{document}
\title{Light-induced electronic polarization in antiferromagnetic Cr$_2$O$_3$}
\author{Xinshu Zhang}
\affiliation{Department of Physics and Astronomy, University of California Los Angeles, Los Angeles, CA 90095, USA}

\author{Tyler Carbin}
\affiliation{Department of Physics and Astronomy, University of California Los Angeles, Los Angeles, CA 90095, USA}

\author{Adrian B. Culver}
\affiliation{Department of Physics and Astronomy, University of California Los Angeles, Los Angeles, CA 90095, USA}
\affiliation{Mani L. Bhaumik Institute for Theoretical Physics, Department of Physics and Astronomy, University of California Los Angeles, Los Angeles, CA 90095, USA}

\author{Kai Du}
\affiliation{Rutgers Center for Emergent Materials, Rutgers University, Piscataway, NJ, USA}

\author{Kefeng Wang}
\affiliation{Rutgers Center for Emergent Materials, Rutgers University, Piscataway, NJ, USA}

\author{Sang-Wook Cheong }
\affiliation{Rutgers Center for Emergent Materials, Rutgers University, Piscataway, NJ, USA}

\author{Rahul Roy}
\affiliation{Department of Physics and Astronomy, University of California Los Angeles, Los Angeles, CA 90095, USA}
\affiliation{Mani L. Bhaumik Institute for Theoretical Physics, Department of Physics and Astronomy, University of California Los Angeles, Los Angeles, CA 90095, USA}

\author{Anshul Kogar}
\email{anshulkogar@physics.ucla.edu}
\affiliation{Department of Physics and Astronomy, University of California Los Angeles, Los Angeles, CA 90095, USA}

\date{\today}

\maketitle

\textbf{In a solid, the electronic subsystem can exhibit incipient order with lower point group symmetry than the crystal lattice. External fields that couple to electronic order parameters have rarely been investigated, however, despite their potential importance to inducing exotic effects. Here, we show that when inversion symmetry is broken by the antiferromagnetic (AFM) order in Cr$_2$O$_3$, transmitting a linearly polarized light pulse through the crystal gives rise to an in-plane rotational symmetry breaking (from $C_3$ to $C_1$) via optical rectification. Using interferometric time-resolved second harmonic generation, we show that the ultrafast timescale of the symmetry reduction is indicative of a purely electronic response; the underlying spin and crystal structures remain unaffected. The symmetry-broken state exhibits a dipole moment, and its polar axis can be controlled with the incident light. 
Our results establish a coherent nonlinear optical protocol by which to break electronic symmetries and produce unconventional electronic effects in solids.
} 


Neumann's principle, formulated in 1885, stipulates that the physical properties of a perfect crystal must possess at least the point group symmetries of the underlying crystal~\cite{symmetry1}. It serves as a cornerstone for interpreting experiments on crystalline matter. In recent years, however, this paradigm has been increasingly questioned. Electronic nematics, 
which are characterized by a lowering of rotational symmetry that is not simply a consequence of reduced lattice symmetry, challenge this framework~\cite{kivelson}. Experiments on electronic nematics illustrate that the symmetry of the electronic subsystem can potentially be decoupled from that of the lattice~\cite{eisenstein, feldman, mackenzie, harter, bozovic, fisher, bozovic2, moll, matsuda}. 

An outstanding related question is whether external fields can be found that break point group symmetries of only the electronic subsystem~\cite{TaAs, TaAsprl}. In the current experimental landscape, it is difficult to disentangle the extent to which the electrons or ions respond to static fields like strain, pressure or electric fields.  
However, ultrafast light pulses can potentially overcome this roadblock. 


One method to isolate the electronic subsystem is to use a Floquet engineering protocol, where an ultrashort light pulse is shone below the electronic gap to avoid absorption and heating~\cite{Floquet,nonlinearFloquet,shirley}. The electrons can then be driven coherently by the light's oscillating fields and potentially be decoupled from the lattice. To date, Floquet engineering has been primarily used in two ways. First, it has been used to break the rotational symmetry of the spin subsystem through the inverse Faraday or inverse Cotton-Mouton effect~\cite{InverseFaraday, InverseFaradayNiO, InverseCMENiO, InverseCM}. Second, the electronic subsystem has been manipulated without a change in symmetry by engineering energy level shifts through the AC Stark and/or Bloch-Siegert effects \cite{MnPS3,WS2,WS22,Yihua,Fahad,mciver}. 
So far, engineering the rotational symmetry of the electronic subsystem
has not been shown; an electronic counterpart to the inverse Faraday effect remains to be discovered.

\begin{figure*}[t]
	\centering
	\includegraphics[width=1.8\columnwidth]{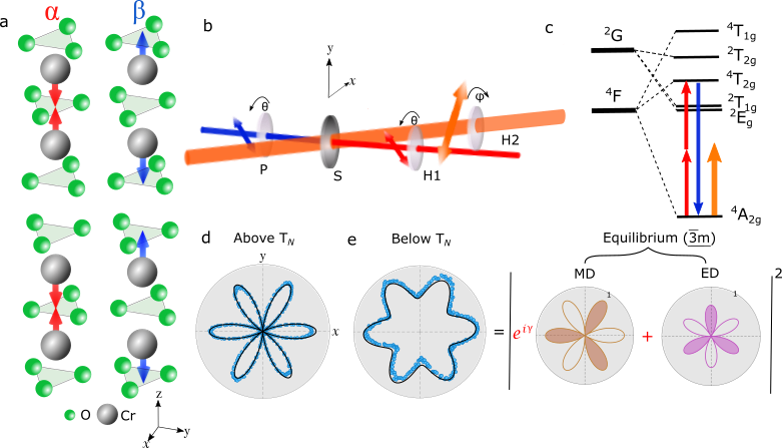}
	\caption{(a) Crystal structure and spin arrangement below $T_N$ in Cr$_2$O$_3$. Two antiferromagnetic (AFM) domain states are possible, which we label $\alpha$ (red) and $\beta$ (blue), where arrows represent the spins on the Cr site. (b) Illustration of our time-resolved second harmonic generation (SHG) experimental setup. H1 is the half wave plate and P is the polarizer to select the polarization of the incident fundamental and output second harmonic light, respectively. The H1 and P rotate together such that the input and output polarization are synchronized at angle $\theta$ relative to $y$-axis. H2 is the half wave plate used to adjust angle of the polarization, $\varphi$, of the pump light relative to the $y$-axis. S denotes the sample. (c) Schematic energy levels of Cr$_2$O$_3$ split by the crystal effective field (CEF). The second harmonic generation process is depicted with the two red arrows (incident photons) and the blue arrow (second harmonic photon). The pump energy is below all the CEF excitations (orange arrow).  (d) Equilibrium rotational anisotropy second harmonic generation (RA-SHG) patterns for the $\beta$ AFM domain state above and (e) below $T_{N}$, at 310~K and 150~K, respectively. Below $T_N$, there are two contributions to the second harmonic signal coming from magnetic dipole radiation (brown) and electric dipole radiation (purple). The shaded regions are contrasted to the transparent regions to indicate where the SHG amplitude changes sign. The magnitudes are normalized to 1.}  
	\label{fig:1}
\end{figure*} 

In this work, by shining an ultrashort light pulse through a noncentrosymmetric antiferromagnet, we show that optical rectification can induce a quasi-DC response that selectively breaks the in-plane rotational symmetry of the electronic subsystem. Optical rectification is a nonlinear process whereby light's oscillating electric field brings about a quasi-DC electric dipole moment~\cite{Boyd, OR, Kaplan}. To leading order, the induced dipole moment is given by:
\begin{equation}
    \boldsymbol{P}(\omega_0) = \boldsymbol{\chi}^{e}_{OR}(\omega_0; \omega_1, -\omega_2) \boldsymbol{E}(\omega_1)\boldsymbol{E}(-\omega_2),
    \label{eq:Rectification}
\end{equation}
where $\boldsymbol{E}(\omega_j)$ ($j$=1, 2) is the electric field vector of the incident light, $\boldsymbol{\chi}^{e}_{OR}(\omega_0; \omega_1, -\omega_2)$ is the second-order nonlinear susceptibility tensor that describes optical rectification, and the frequency of the rectified response is given by $\omega_0 = \omega_1 - \omega_2\approx 0$ with $\omega_1$ and $\omega_2$ within the energy spread of the light pulse. Crucially, when light pulses shorter than the typical structural response timescales are used to generate a rectified dipole moment, one can distinguish between a structural and an electronic response via the system's relaxation dynamics. 

To observe the induced symmetry breaking, we probe the system with rotational anisotropy second harmonic generation (RA-SHG), a technique sensitive to electronic symmetries \cite{SHGastool}. In our experimental configuration, both the incident fundamental and detected second harmonic light beams are polarized along the same direction (Fig.~\ref{fig:1}(b)). As a model compound, we select the prototypical linear magnetoelectric Cr$_2$O$_3$ because, as we describe below, interference of the electric and magnetic dipole second harmonic radiation allows for high-sensitivity detection of broken point group symmetries \cite{FiebigSHG, topography, timeCr2O3,timeCr2O32,timeCr2O3wall}. 

At equilibrium, Cr$_2$O$_3$ undergoes an antiferromagntic transition at $T_{N}\approx$~307~K. Above $T_N$, Cr$_2$O$_3$ possesses $\bar{3}m$ (D$_{3d}$) point group symmetry. Electric dipole SHG is forbidden in this state due to the presence of inversion symmetry. However, when the second harmonic energy is tuned to a Cr $d$-$d$ electronic transition ($^{4}\hspace{-0.1cm}A_{2g} \hspace{0.1cm} (t_{2g})^3 \hspace{-0.15cm}\rightarrow^4\hspace{-0.1cm}T_{2g} \hspace{0.1cm} (t_{2g})^2e_g$) at 2.1~eV (590~nm), we observe resonant magnetic dipole SHG (Fig.~\ref{fig:1}(c)-(d))~\cite{FiebigSHG}. As shown in Fig.~\ref{fig:1}(d), the RA pattern is consistent with the threefold symmetry of the crystal when the probe light propagates along the out-of-plane direction (and is polarized in-plane). 
\begin{figure*}[t]
	\centering
	\includegraphics[width=1.73\columnwidth]{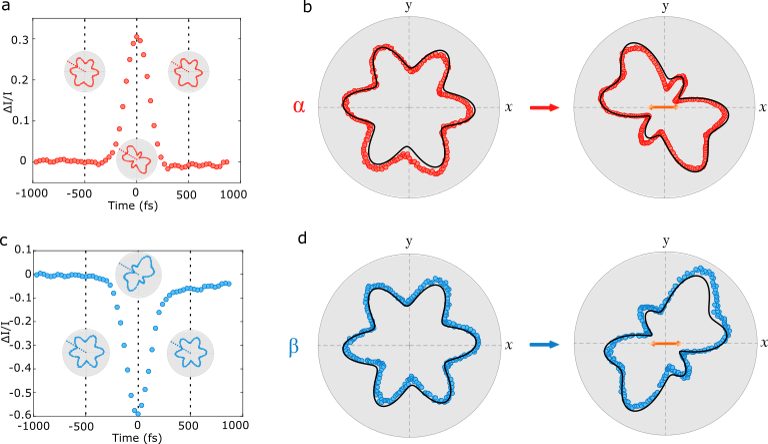}
	\caption{(a) The normalized change to the SHG intensity at $T=150$~K in the $\alpha$ domain state as a function of time delay for pump light aligned along $x$-axis and probe fixed to $\theta \sim 60^{\circ}$. The fluence of the pump light is 20~mJ/cm$^2$. Insets show three representative RA patterns at $\pm$500~fs and 0~fs. The red dashed line in RA pattern indicates the chosen direction of the probe polarization for the time scans. (b) Change from  sixfold RA-SHG pattern (left panel; unpumped) to twofold pattern (right panel; pumped) upon pumping the $\alpha$ domain state. The orange double headed arrow indicates the orientation of the pump polarization. (c) Same as (a), but in the $\beta$ domain state. (d) Same as (b), but in the  $\beta$  domain state. The solid black line in (d) is the fit using the function described in main text. Note that the black line in (b) is not a fit. It is obtained by  flipping the sign of electric dipole contribution of the fit in (d) while keeping the magnetic dipole contribution fixed.}  
	\label{fig:2}
\end{figure*} 

Below $T_{N}$, Cr$_2$O$_3$ orders antiferromagnetically with the four Cr spins in the unit cell alternating in an up and down sequence along the rhombohedral optical axis (Fig.~\ref{fig:1}(a)). This spin structure breaks inversion symmetry, and the magnetic point group becomes $\underline{\bar{3}m}$~\cite{Birss}. Electric dipole SHG is then allowed (through the spin-orbit interaction) and interferes with the pre-existing magnetic dipole signal~\cite{ME0, ME1, ME2}. 
Due to the presence of both magnetic and electric dipole signals, the nodes present in the RA pattern above $T_N$ are lifted below $T_N$ (Fig.~\ref{fig:1}(e)). That the nodes are lifted implies a relative phase (not zero or 180$^\circ$) between the magnetic and electric dipole SHG amplitudes at the probed wavelength. The RA pattern below $T_N$ can be fit with a simple function that includes electric and magnetic dipole radiation (Fig.~\ref{fig:1}(e))~\cite{FiebigSHG}:
\begin{equation}
    I(2\omega_{pr}) \propto |e^{i\gamma} \chi^m\textrm{sin}(3\theta) \pm \chi^e\textrm{cos}(3\theta)|^2 \\
    \label{eq:RA-SHG}
\end{equation}
where $\chi^{e/m}$ is the real-valued in-plane electric/magnetic dipole second harmonic susceptibility, the $\pm$ depends on the AFM domain state and $\gamma$ denotes the relative phase between magnetic and electric dipole second harmonic radiation (see Supplementary Note \RNum{1}). Here, $\omega_{pr}$ is the frequency of the probe light and $\theta$ represents the polarization angle of the incident and detected light with respect to the sample's $y$-axis (Fig.~\ref{fig:1}~(b)). Below $T_N$, the electric dipole SHG susceptibility, $\chi^e$, is proportional to the AFM order parameter, \textbf{L}~\cite{FiebigSHG}. Thus, $\chi^e$ differs in sign between the two AFM domain states, which we denote $\alpha$ and $\beta$ (Fig.~\ref{fig:1}(a))~\cite{FiebigSHG,topography}. In Eq.~\ref{eq:RA-SHG}, it is also important to note that $\gamma$ depends sensitively on the second harmonic energy (Supplementary Note \RNum{6}). At the 2.1~eV second harmonic energy used here, $\gamma \approx$ 85$^\circ$ and the two domain states exhibit almost identical RA patterns (left panels of Fig.~\ref{fig:2}(b) and (d))~\cite{Cr2O3book}. This equivalence can be understood by noting that cross terms in Eq.~\ref{eq:RA-SHG} vanish when $\gamma = 90^\circ$ which eliminates the contrast between domain states. The interference between the electric and magnetic dipole SHG equips the crystal with an inherent phase sensitivity, which is key to observing the large symmetry-breaking response that we now demonstrate (Supplementary Note \RNum{6}).

We now move on to pump the Cr$_2$O$_3$ crystal with 1.2~eV (1030~nm) light pulses with fluences up to 20~mJ/cm$^2$. The pump polarization is aligned along the $x$-axis, as indicated with the double headed arrows in the right panels of Fig.~\ref{fig:2}(b) and (d). The wavelength of the pump light lies in the transparency window of the crystal and away from electronic resonances to avoid significant absorption and heating (orange arrow in Fig.~\ref{fig:1}(c)). As the pump is transmitted through the crystal, we observe a drastic symmetry change in the RA pattern (right panels of Fig.~\ref{fig:2}(b) and (d)). Additionally, the two domain states, which are almost indistinguishable with RA-SHG at equilibrium, exhibit differing responses to the pump. (The two domain states are distinguishable at equilibrium with circularly polarized SHG~\cite{FiebigSHG, topography}). 
This domain state-dependent symmetry reduction is not observed above $T_N$ or when the pump light is circularly polarized; instead both broken inversion symmetry (due to the antiferromagnetism) in addition to a well-defined pump polarization axis are necessary to observe this light-induced symmetry-breaking effect in Cr$_2$O$_3$ (Supplementary Notes \RNum{3} and \RNum{5}).

\begin{figure*}[t]
	\centering
	\includegraphics[width=1.76\columnwidth]{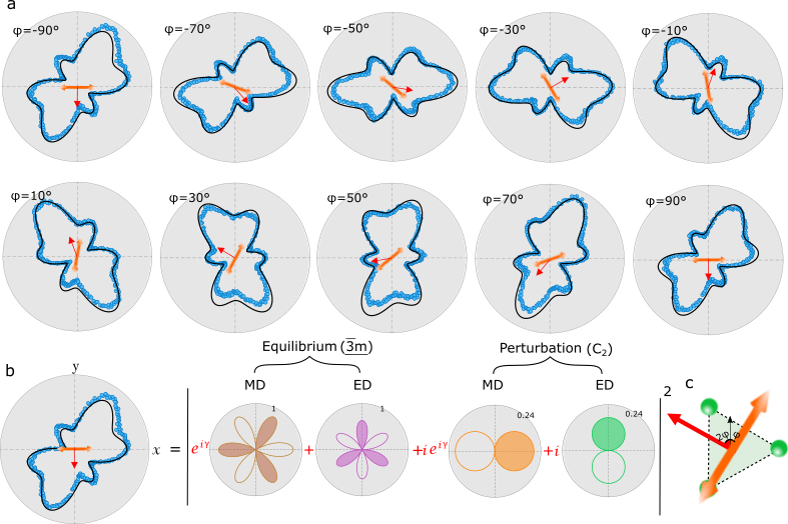}
	\caption{(a) Evolution of RA pattern at $T=150$~K in the $\beta$ domain state as the pump polarization is varied. Starting along $x$-axis, the polarization is rotated clockwise in 20$^{\circ}$ steps (orange double headed arrow). The pump polarization is defined with respect to $y$-axis. The orientation of the light-induced electronic dipole is shown with the red single-headed arrow. In these plots, only the pattern with $\varphi=-90^\circ$ is fit, while the rest of the black curves are obtained using Eq.~\ref{eq:PerturbationPhase} of the main text. (b) The four contributions to the RA pattern consist of the equilibrium magnetic and electric dipole amplitudes in addition to the pump-induced perturbations to the magnetic and electric dipole signals. The perturbations have a polar nature as indicated by the shaded and transparent regions. These contributions are shown in brown, purple, orange and green. (c) A schematic showing the orientation of the induced electronic dipole (red) and the pump polarization (orange).  }  
	\label{fig:3}
\end{figure*} 
Importantly, we observe this effect only when the pump and probe pulses are temporally overlapped; the timescale characterizing this transient symmetry breaking is on the order of the laser pulse width, $\sim$180~fs (Fig.~\ref{fig:2}(a) and (c)). (The slight asymmetry in the background levels of the time traces is attributed to two-photon absorption and is unrelated to the symmetry-breaking.) In the insets of Fig.~\ref{fig:2}(a) and (c), we show RA patterns 500~fs before and after the pump pulse propagates through the probed region as well as the pattern when the pump and probe pulses are perfectly overlapped (the latter are also shown in the right panels of Fig.~\ref{fig:2}(b) and (d)). Notably, there is not a measurable relaxation timescale for the symmetry change (apart from the pulse width), which suggests that neither the structural nor spin degrees of freedom bring about this reduction in rotational symmetry (Supplementary Note~\RNum{7}). In contrast to a previous similar study on Cr$_2$O$_3$~\cite{timeCr2O3wall}, here, the electronic charge degree of freedom is solely involved (Supplementary Note~\RNum{12}).

\begin{figure*}[t]
	\centering
	\includegraphics[width=1.8\columnwidth]{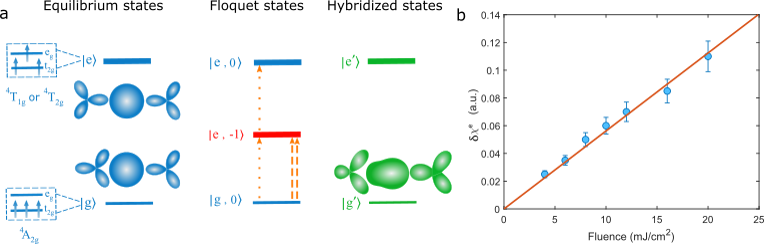}
\caption{(a) Energy levels and associated electronic probability densities in the $x-y$ plane demonstrating how the light-induced electric dipole arises via optical rectification. At equilibrium, the ground state $| g\rangle$ ($^{4}\hspace{-0.1cm}A_{2g}$) in  Cr$_2$O$_3$ comprises three electrons in $t_{2g}$ state. (The symmetry of the states is discussed in Supplementary Note \RNum{8}). The relevant excited state manifold $| e\rangle$ comprises both the $^{4}T_{2g}$ and $^{4}T_{1g}$ states, which consist of two $t_{2g}$ electrons and one $e_{g}$ electron (we neglect the $\sim$0.6~eV splitting between the $^4T_{2g}$ and $^4T_{1g}$ states for clarity). Representative electron probability densities are also shown. In the middle panel, we show that when applying the periodic drive a single Floquet state, $| e,- 1\rangle$, is generated within the rotating wave approximation. 
There are two possible channels for generating a static electric dipole, which are indicated by the orange arrows (discussed further in the main text). In the right panel, representative electron probability densities for the first dipole channel are shown, and a dipole can clearly be observed.
(b) The fluence dependence of $\delta \chi^{e}$ (proportional to light-induced electric dipole, $\boldsymbol{P}(\omega_0)$) at $T=150$~K.}  
	\label{fig:4}
\end{figure*} 

The transient electronic symmetry reduction originates from a light-induced electronic dipole moment (via optical rectification). 
Under this interpretation, the electric/magnetic dipole susceptibility tensor describing SHG can be expanded to first order in the induced moment:
\begin{equation}
    \begin{split}
    \boldsymbol{\chi}^{e/m}_{SHG}(\textbf{\textit{P}}(\omega_0)) = \left.\boldsymbol{\chi}^{e/m}_{SHG}\right|_{\textbf{\textit{P}}=0} + \underbrace{\left.\frac{\partial\boldsymbol{\chi}_{SHG}^{e/m}}{\partial\textbf{\textit{P}}(\omega_0)}\right|_{\textbf{\textit{P}}=0}\hspace{-0.5cm}\textbf{\textit{P}}(\omega_0)}_{\equiv\delta\boldsymbol{\chi}^{e/m}_{SHG}}.
    \label{eq:cascade}
    \end{split}
\end{equation}
The rectified dipole moment, $\textbf{\textit{P}}({\omega_0})$, can be calculated based on the magnetic symmetry of the crystal for which we obtain the following expression (see Supplementary Note \RNum{2}A): 
\begin{equation}
 \textbf{\textit{P}}({\omega_0})=  \chi^{e}_{OR} I_{pump} (\textrm{sin}(-2\varphi) \hat{\textbf{x}}+ \textrm{cos}(-2\varphi)  \hat{\textbf{y}}),
\label{eq:DipoleMoment}
\end{equation} 
where $\chi^e_{OR}$ is the in-plane susceptibility associated with optical rectification, $I_{pump}$ is the intensity of the pump beam, and $\varphi$ represents the pump polarization angle with respect to the sample's $y$-axis.
By combining the effective susceptibility tensor (Eq.~\ref{eq:cascade}) with the expression for the pump induced dipole moment (Eq.~\ref{eq:DipoleMoment}), we can derive the full expression for the SHG intensity as function of the pump polarization angle for the two domain states (Supplementary Note \RNum{2}B):
\begin{equation}
\begin{split}
I(2\omega_{pr},  \hspace{0.05cm}&\varphi) \propto |e^{i\gamma}\chi^m\textrm{sin}(3\theta) \pm \chi^e\textrm{cos}(3\theta) \\
&+ie^{i\gamma}\delta\chi^m\textrm{sin}(\theta-2\varphi) \pm i\delta\chi^e\textrm{cos}(\theta-2\varphi)|^2.
\label{eq:PerturbationPhase}
\end{split}
\end{equation}
Compared to the equilibrium case, only two additional independent terms are permitted to model the RA pattern: perturbations to the in-plane electric and magnetic dipole susceptibilities, $\delta\chi^e$ and $\delta\chi^m$, respectively. 
In Fig.~\ref{fig:3}(b), we demonstrate that Eq.~\ref{eq:PerturbationPhase} yields an excellent fit to the RA pattern when the pump polarization is along $x$-axis (i.e.  $\varphi=-90^\circ$). (In Supplementary Note \RNum{2}C, we explain how the combined parity and time-reversal symmetry accounts for the factor of $i$ multiplying the perturbative terms in Eq.~\ref{eq:PerturbationPhase}).

Our assignment of a light-induced electronic dipole moment is corroborated by two non-trivial predictions of our fit model. First, as the pump polarization angle, $\varphi$, is varied, the induced dipole moment is correspondingly rotated by an angle $-2\varphi$ (Eq.~\ref{eq:DipoleMoment} and \ref{eq:PerturbationPhase}). In Fig.~\ref{fig:3}(c), we illustrate this concept visually with the orange and red arrows which represent the orientation of the pump polarization and induced dipole moment, respectively. Once the RA pattern at $\varphi=-90^\circ$ is fit (Fig.~\ref{fig:3}(b)), the dependence on the pump polarization angle, $\varphi$, is completely determined by Eq.~\ref{eq:PerturbationPhase}. In Fig.~\ref{fig:3}(a), we show the dependence of the RA pattern on the pump polarization angle $\varphi$, while the corresponding results generated using Eq.~\ref{eq:PerturbationPhase} are shown in black. Even though the RA patterns change shape as the pump polarization is tuned, the black curves nonetheless show remarkable agreement with the data. We emphasize that these are not fits to the individual RA patterns; only the $\varphi =-90^\circ$ pattern is fit and the remaining black lines are generated with Eq.~\ref{eq:PerturbationPhase}, where only a single parameter, $\varphi$, is varied. Second, by choosing the opposite sign of $\chi^e$ and $\delta\chi^e$ in Eq.~\ref{eq:PerturbationPhase} (terms proportional to the order parameter), we obtain excellent agreement to the RA pattern of the opposite antiferromagnetic domain state. Again, the black line in the right panel of Fig~\ref{fig:2}(b) is not a fit, but is generated from this procedure. This scheme implies that the two AFM domain states exhibit opposite induced dipole moments.  


The model itself does not explain how the symmetry is broken from a microscopic standpoint, however. To understand how this is accomplished, we appeal to the useful picture provided by Floquet theory~\cite{shirley, nonlinearFloquet}. In the leftmost panel of Fig.~\ref{fig:4}(a), we schematically illustrate the energy levels of the Cr atoms in the crystal field environment with the ground state  $\ket{g}$ and the excited states $\ket{e}$, and show examples of the corresponding in-plane probability densities.  When a periodic potential is applied to the system, a series of Floquet sidebands or ``dressed states" emerge, which we label $\ket{j, n}$, where $j$ labels the equilibrium state from which the $n^{th}$ Floquet sideband derives. In the middle panel of Fig.~\ref{fig:4}(a), we make the rotating wave approximation and only show a single dressed state of the excited state manifold, $\ket{e, -1}$. 
In this effectively time-independent scheme, a superposition state, $\ket{g'} = \ket{g,0} + \lambda\ket{e,-1} + \lambda^2\ket{e,0}$ forms, where the numerator of $\lambda$ is given by matrix elements of the form $\bra{j',n\pm1}\boldsymbol{d}\cdot\boldsymbol{E}\ket{j,n}$, and the power of $\lambda$ indicates a first or second order perturbative correction to the ground state~\cite{shirley}. Here, $\boldsymbol{d}$ is the dipole operator and $\boldsymbol{E}$ is the time-independent electric field vector. Such matrix elements sensitively depend on the polarization of the incoming light. 

The new hybridized state, $\ket{g'}$, is capable of exhibiting a static dipole moment, which can be seen by calculating the expectation value $\bra{g'}\boldsymbol{d}\ket{g'}$ (Supplementary Note \RNum{9}). With the orange arrows in the middle panel of Fig.~\ref{fig:4}(a), we illustrate the two pathways through which the dipole can develop. The first path is shown with the dotted arrows and indicates that a dipole can arise due to terms of the form $\lambda^2\bra{g,0}\boldsymbol{d}\ket{e,0}\sim|\boldsymbol{E}|^2$. A second pathway, shown with the dashed arrows, yields terms of the form $\lambda^2\bra{e,-1}\boldsymbol{d}\ket{e,-1}\sim |\boldsymbol{E}|^2$. Selection rules associated with such matrix elements have been previously been calculated for Cr$_2$O$_3$ in Refs.~\cite{Cr2O3book, ME0, ME1, ME2}. We use the matrix elements therein to sketch representative probability densities for the first kind of process, which we show in the rightmost panel of Fig.~\ref{fig:4}(a). Clearly, the in-plane threefold symmetry of the crystal is broken and a finite dipole moment can be observed. (Supplementary Note \RNum{8} clarifies the symmetries of the sketched probability densities. Further details of the calculation using time-dependent perturbation theory and Floquet theory are presented in Supplementary Note \RNum{9}.)

It is important to note that both pathways give rise to terms that scale with $\lambda^2 \sim |\boldsymbol{E}|^2$. Such a relation implies that the lowest order static electric dipole moment scales with $|\boldsymbol{E}|^2$ (i.e. fluence), which is confirmed experimentally in Fig.~\ref{fig:4}(b). This scaling is consistent with the second order nonlinear optical process producing a field conjugate to the electronic dipole moment.



In conclusion, our experimental observations and simple theoretical account demonstrate how Floquet engineering can be used to manipulate the symmetry of electronic orbitals. The macroscopic symmetry of the electronic subsystem subsequently decouples from that of the lattice on an ultrafast timescale to break in-plane rotational symmetry and to produce a purely electronic dipole moment. 
This experiment builds on previous work showing that purely electronic macroscopic effects in crystals can be brought about through the coherent nonlinear interaction of light with matter~\cite{InverseFaraday, InverseFaradayNiO, InverseCMENiO, InverseCM, mciver, MnPS3, WS2, WS22, Yihua, Fahad}. 
Our work paves the way towards quantifying the electronic contribution to the ferroelectric effect, engineering electronic phases with light, and manipulating magnetism via an electronic magnetoelectric effect in candidate materials \cite{contribution1,contribution2,magnetism1, type2MEa,type2MEb,reviewME1,reviewME2}.

\footnotesize
\vspace{-0.75em}
 \section{Acknowledgements:}
We thank Mengxing Ye, Honglie Ning, Carina Belvin and Wesley Campbell for helpful conversations related to this work.
Research at UCLA was supported by the U.S. Department of Energy (DOE), Office of Science, Office of Basic Energy Sciences under Award No. DE-SC0023017 (experiment and theory). The work at Rutgers was supported by W. M. Keck Foundation (materials synthesis).  A.B.C. and R.R. acknowledge financial support from the University of California Laboratory Fees Research Program funded by the UC Office of the President (UCOP), grant number LFR-20-653926.  A.B.C acknowledges financial support from the Joseph P. Rudnick Prize Postdoctoral Fellowship (UCLA).

\section{Author contributions:  }
X.Z. and T.C. built the SHG setup and performed the time-resolved SHG experiments under the supervision of A.K. X.Z. analysed the data under the supervision of A.K. K.D. and K.W. grew the single crystals under the supervision of S.-W.C. Theoretical calculations were carried out by A.B.C. with input from R.R., X.Z and A.K. The manuscript was written by X.Z., A.B.C. and A.K. with input from all authors.
\vspace{-0.75em}

\section{Competing interests:  }

The authors declare no competing interests.

\section{Methods }
\vspace{-1em}
\subsection{Sample synthesis}
Cr$_2$O$_3$ single crystals were grown using a laser diode heated floating zone (LFZ) technique. Cr$_2$O$_3$ powders (Alfa Aesar, $99.99\%$) were pressed into 3 mm diameter rods under 8000 PSI hydrostatic pressure. The compressed rod was sintered at 1600$^{\circ}$C in a box furnace for 10~hours. The crystals were grown with growth speed of 2 to 4 mm/h in oxygen flow of 0.1 l/min, and counter rotation of the feed and seed rods at 15 and 15 rpm, respectively.

\subsection{Experimental details}
The regeneratively amplified laser used in our experiment is based on a Yb:KGW gain medium that outputs a power of 10~W. The laser pulses have a Gaussian-like profile with an approximately 180~fs pulse duration and a 1030~nm central wavelength. In our experiment, we used a laser pulse repetition rate of 5~kHz. The fundamental output of the laser at 1030~nm was used as the pump pulse, which was focused obliquely on the sample at a 10 degree angle of incidence. The pump laser spot size was $\sim$ 500~$\mu$m and the maximum fluence was $\sim$ 20 $\mathrm{mJ}/\mathrm{cm}^{2}$. The probe pulse was generated from an optical parametric amplifier with tunable wavelength, which we use for the second harmonic spectroscopy between 900-1200~nm. The probe pulse was focused normally on the sample with a 100~$\mu$m spot size, and the probe fluence was $\sim$~2~$\mathrm{mJ}/\mathrm{cm}^{2}$. Detection of the second harmonic light was conducted with a commercial photo-multiplier tube. The sample was cooled to 150~K with a standard optical cryostat with fused silica windows to prevent distortions to the light polarization.

 \section{Data availability}
 \vspace{-1em}
 The data that supports the findings of this study are present in the paper and/or in the supplementary information, and are deposited in the Zenodo repository.  Additional data related to the paper is available from the corresponding authors upon reasonable request.

\normalsize


\begin{thebibliography}{1}
\bibitem{symmetry1}    Powell, R.  Symmetry, group theory, and the physical properties of crystals \textit{Springer 
}  (2010).
\bibitem{kivelson} Kivelson, S.A., Fradkin, E. $\&$ Emery, V. Electronic liquid-crystal phases of a doped Mott insulator. \textit{Nature} \textbf{393}, 550-553 (1998).
\bibitem{eisenstein} Lilly, M.P. et al. Evidence for an anisotropic state of two-dimensional electrons in high Landau levels. \textit{Phys. Rev. Lett.} \textbf{82}, 394 (1999).
\bibitem{feldman} Feldman, B.E. et al. Observation of a nematic quantum Hall liquid on the surface of bismuth. \textit{Science} \textbf{354}, 316-321 (2016).
\bibitem{mackenzie} Borzi, R. et. al. Formation of a nematic fluid at high fields in Sr$_3$Ru$_2$O$_7$ at high magnetic fields. \textit{Science} \textbf{315}, 214-217 (2007).
\bibitem{harter} Harter, J. et al. A parity breaking electronic nematic phase transition in the spin orbit coupled metal Cd$_2$Re$_2$O$_7$ \textit{Science} \textbf{356}, 295-299 (2017).
\bibitem{bozovic} Wu, J. et al. Spontaneous breaking of rotational symmetry in copper oxide superconductors. \textit{Nature} \textbf{547}, 432-435 (2017).
\bibitem{fisher} Chu, J. et al. Divergent nematic susceptibility in an iron arsenide superconductor. \textit{Science} \textbf{337}, 710-712 (2012).
\bibitem{bozovic2} Wu, J. et al. Electronic nematicity in Sr$_2$RuO$_4$. \textit{Proc. Natl. Acad. Sci.} \textbf{117}, 10654-10659 (2020).
\bibitem{moll} Ronning, F. et al. Electronic in-plane symmetry breaking at field-tuned quantum criticality in CeRhIn$_5$. \textit{Nature} \textbf{548}, 313-317 (2017).
\bibitem{matsuda} Okazaki, R. et al. Rotational symmetry breaking in the hidden-order phase of URu$_2$Si$_2$. \textit{Science} \textbf{331}, 439-442
(2011)
\bibitem{TaAs}   Sirica, N. et al.  Photocurrent-driven transient symmetry breaking in the Weyl semimetal TaAs. \textit{Nat. Mater.} \textbf{ 21, } 62–66 (2022).
\bibitem{TaAsprl}   Sirica, N. et al. Tracking Ultrafast Photocurrents in the Weyl Semimetal TaAs Using THz Emission Spectroscopy. \textit{Phys. Rev. Lett. } \textbf{ 122, } 197401 (2019).




\bibitem{shirley}  Shirley, J. Solution of the Schrodinger equation with a Hamiltonian periodic in time \textit{Phys. Rev} \textbf{138,  } 4B (1965).
\bibitem{Floquet}  Oka, T. $\&$ Kitamura, S.  Floquet Engineering of quantum materials. \textit{Annu. Rev. Condens. Matter Phys.} \textbf{10,}  387-408 (2019).
\bibitem{nonlinearFloquet}  Morimoto, T.  $\&$ Nagaosa, N. Topological nature of nonlinear optical effects in solids. \textit{Sci. Adv.} \textbf{2,  } 5 (2016).
\bibitem{InverseFaraday}    Kimel, A. et al. Ultrafast non-thermal control of magnetization by instantaneous photomagnetic pulses. \textit{Nature} \textbf{435,   }655–657 (2005).
\bibitem{InverseFaradayNiO} Satoh, T. et al. Spin oscillations in antiferromagnetic NiO triggered by circularly polarized light. \textit{Phs. Rev. Lett.} \textbf{105}, 077402 (2010).
\bibitem{InverseCMENiO} Tzschaschel, C. et al. Ultrafast optical excitation of coherent magnons in antiferromagnetic NiO. \textit{Phs. Rev. B} \textbf{95}, 174407 (2017).
\bibitem{InverseCM} Pershan, P.S., van der Ziel, J.P. $\&$ Malmstrom, L.D. \textit{Phys. Rev.} \textbf{143}, 574-583 (1966).
\bibitem{MnPS3}  Shan, J. et al. Giant modulation of optical nonlinearity by Floquet engineering \textit{Nature} \textbf{ 600, } 235–239 (2021).
\bibitem{WS2}   Sie, E. et al. Valley-selective optical Stark effect in monolayer WS$_2$. \textit{Nat. Mater.} \textbf{14, } 290–294 (2015).
\bibitem{WS22}   Sie, E. et al. Large, valley-exclusive Bloch-Siegert shift in monolayer WS$_2$. \textit{Science} \textbf{355,  } 6329 (2015).

\bibitem{Yihua}  Wang, Y. et al. Observation of Floquet-Bloch States on the Surface of a Topological Insulator \textit{Science} \textbf{342,    } 453–457 (2013).
\bibitem{Fahad}  Mahmood, F. et al. Selective scattering between Floquet–Bloch and Volkov states in a topological insulator. \textit{Nat. Phys.} \textbf{12,    } 306–310 (2016).
\bibitem{mciver} McIver, J.W. et. al. Light-induced anomalous Hall effect in graphene. \textit{Nat. Phys.}, \textbf{16} 38-41 (2020). 

\bibitem{Boyd}   Boyd, R.  Nonlinear optics. \textit{Academic Press } (2020).

\bibitem{OR}  Bass, M. et al. Optical Rectification. \textit{Phys. Rev. Lett.} \textbf{9}, 446 (1962).

\bibitem{Kaplan} Kaplan, D., Holder, T. and Yan, B. Nonvanishing subgap photocurrent as a probe of lifetime effects. \textit{Phys. Rev. Lett.} \textbf{125}, 227401 (2020).


\bibitem{SHGastool}  Fiebig, M.,  Pavlov, V.  $\&$ Pisarev, R.   Second-harmonic generation as a tool for studying electronic and magnetic structures of crystals: review \textit{J. Opt. Soc. Am. B} \textbf{22,    } 1, 96-118 (2005).
\bibitem{FiebigSHG}  Fiebig, M. et al.   Second Harmonic Generation and Magnetic-Dipole —Electric-Dipole Interference in Antiferromagnetic Cr$_2$O$_3$ \textit{Phys. Rev. Lett.} \textbf{73}, 2127 (1994).
\bibitem{topography}    Fiebig, M.,Fröhlich, D. $\&$   Sluyterman, G. Domain topography of antiferromagnetic Cr$_2$O$_3$  by second‐harmonic generation. \textit{Appl. Phys. Lett.} \textbf{ 66},  2906 (1995).
\bibitem{timeCr2O3}  Satoh, T. et al.  Ultrafast spin and lattice dynamics in antiferromagnetic Cr$_2$O$_3$ \textit{Phys. Rev. B} \textbf{ 75}, 155406 (2007).
\bibitem{timeCr2O32}  Satoh, T. et al.  Time-resolved demagnetization in  Cr$_2$O$_3$ by phase sensitive second harmonic generation \textit{Phys. Rev. B} \textbf{310}, 1604-1606 (2007).
\bibitem{timeCr2O3wall}  Sala, V. et al.  Resonant optical control of the structural distortions that drive ultrafast demagnetization in  Cr$_2$O$_3$  \textit{Phys. Rev. B} \textbf{94}, 014430 (2015).

\bibitem{Birss} Birss, R. Symmetry and Magnetism. \textit{North Holland},  (1966).

\bibitem{ME0}   Muthukumar, V.,   Valentí, R.  $\&$  Gros, C.  Microscopic Model of Nonreciprocal Optical Effects in  Cr$_2$O$_3$  \textit{Phys. Rev. Lett.} \textbf{ 75, 2766 } 2766  (1995).
\bibitem{ME1}   Muto, M. et al.   Magnetoelectric and second-harmonic spectra in antiferromagnetic  Cr$_2$O$_3$ \textit{Phys. Rev. B  } \textbf{57,  } 9586 (1998).
\bibitem{ME2}    Muthukumar, V.,   Valentí, R.  $\&$  Gros, C.   Theory of nonreciprocal optical effects in antiferromagnets: The case of  Cr$_2$O$_3$ \textit{Phys. Rev. B  } \textbf{54,   } 433 (1996).
\bibitem{Cr2O3book}  Tanabe, Y., Fiebig, M. $\&$  Hanamura, E. in Magneto-optics eds. Sugano, S. $\&$ Kojima, N.  \textit{Springer} \textbf{128}, 107  (1999).

\bibitem{contribution1}   Malashevich, A. et al.  Full magnetoelectric response of Cr$_2$O$_3$ from first principles \textit{Phys. Rev. B } \textbf{86,}   094430 (2012).
\bibitem{contribution2}   Bousquet, E.,  Spaldin, N. $\&$  Delaney, K.  Unexpectedly Large Electronic Contribution to Linear Magnetoelectricity \textit{Phys. Rev. Lett. } \textbf{106,}   107202 (2011).
\bibitem{magnetism1}   Kirilyuk, A.,  Kimel, A. $\&$  Rasing, T. Ultrafast optical manipulation of magnetic order  \textit{Rev. Mod. Phys.} \textbf{82,    } 2731 (2010).
\bibitem{type2MEa}  Khomskii, D. Classifying multiferroics: mechanisms and effects. \textit{Physics } \textbf{2,    } 20 (2009).
\bibitem{type2MEb}   Tokura, Y., Seki, S. $\&$ Nagaosa, N. Multiferroics of spin origin. \textit{Rep. Prog. Phys.} \textbf{77,    } 076501 (2014)
\bibitem{reviewME1}   Spaldin, N.  $\&$  Ramesh, R.  Advances in magnetoelectric multiferroics \textit{Nat. Mater.} \textbf{18,}   203–212 (2019).
\bibitem{reviewME2}  Spaldin, N.  $\&$  Fiebig, M.   The Renaissance of Magnetoelectric Multiferroics \textit{Science } \textbf{309,}   391-392 (2005).





\end{thebibliography}
\end{document}



\newcommand{\pd}{\partial}
\newcommand{\beq}{\begin{equation}}
\newcommand{\eeq}{\end{equation}}
\newcommand{\bseq}{\begin{subequations}}
\newcommand{\eseq}{\end{subequations}}
\newcommand{\bpmat}{\begin{pmatrix}}
\newcommand{\epmat}{\end{pmatrix}}
\title{Supplementary Information for Light-induced electronic polarization in antiferromagnetic Cr$_2$O$_3$}
\author{Xinshu Zhang}

\affiliation{Department of Physics and Astronomy, University of California Los Angeles, Los Angeles, CA 90095, USA}

\author{Tyler Carbin}
\affiliation{Department of Physics and Astronomy, University of California Los Angeles, Los Angeles, CA 90095, USA}

\author{Adrian B. Culver}
\affiliation{Department of Physics and Astronomy, University of California Los Angeles, Los Angeles, CA 90095, USA}
\affiliation{Mani L. Bhaumik Institute for Theoretical Physics, Department of Physics and Astronomy, University of California Los Angeles, Los Angeles, CA 90095, USA}

\author{Kai Du}
\affiliation{Rutgers Center for Emergent Materials, Rutgers University, Piscataway, NJ, USA}

\author{Kefeng Wang}
\affiliation{Rutgers Center for Emergent Materials, Rutgers University, Piscataway, NJ, USA}

\author{Sang-Wook Cheong }
\affiliation{Rutgers Center for Emergent Materials, Rutgers University, Piscataway, NJ, USA}
\author{Rahul Roy}
\affiliation{Department of Physics and Astronomy, University of California Los Angeles, Los Angeles, CA 90095, USA}
\author{Anshul Kogar}
\email{anshulkogar@physics.ucla.edu}
\affiliation{Department of Physics and Astronomy, University of California Los Angeles, Los Angeles, CA 90095, USA}

\date{\today}

\maketitle


\section{Analysis of equilibrium SHG }
\subsection{Rotational anisotropy SHG}
Radiation can come from various sources, including electric dipoles, magnetic dipoles, electric quadrupoles, and higher order radiation sources. In Cr$_2$O$_3$, magnetic and electric dipole radiation are known to dominate the response. The total source term entering Maxwell's equations is given by 
\begin{equation}
    \boldsymbol{S}=\left(\nabla \times \frac{\partial  \boldsymbol{M}}{\partial t}+\frac{\partial^{2}  \boldsymbol{P}}{\partial t^{2}}\right).
\end{equation}
The second harmonic intensity measured by the detector in our experiment is $I \propto |\textbf{e}(2\omega)\cdot \boldsymbol{S} |^2$, where  $\textbf{e}(2\omega)$ is the unit vector of the output polarizer that selects second harmonic light of a particular polarization. The magnetization at the second harmonic frequency is allowed at all temperatures and has the form of $\boldsymbol{M}(2\omega)=\boldsymbol{\chi}^{m}\boldsymbol{E}(\omega)\boldsymbol{E}(\omega)$, where $\boldsymbol{\chi}^{m}$ is the second order nonlinear magnetic susceptibility tensor. The electric polarization at the second harmonic frequency is allowed only below $T_{N}$ and is given by $\boldsymbol{P}(2\omega)=\boldsymbol{\chi}^{e}\boldsymbol{E}(\omega)\boldsymbol{E}(\omega)$.  Below $T_N$, we can write the SHG intensity as 
$I \propto |M(\underline{\bar{3}m})+P(\underline{\bar{3}m})|^2$, where $M(\underline{\bar{3}m})$ and $P(\underline{\bar{3}m})$ represent the magnetic and electric dipole contributions to SHG signal at a particular polarization angle. Additionally, $\underline{\bar{3}m}$ is the magnetic point group of the crystal. For light along $z$-direction (optical axis of the crystal), the electric and magnetic susceptibility tensors have the same independent tensor elements with the relation $\chi_{yyy}$=$-\chi_{yxx}$=$-\chi_{xyx}$=$-\chi_{xxy}$=$\chi^{e/m}$. By taking the polarization of both the incident fundamental light and the outgoing SHG light into account, we obtain the magnetic dipole contribution, $M(\underline{\bar{3}m}) \propto \chi^{m}\textrm{sin}(3\theta)$, and the electric dipole contribution, $P(\underline{\bar{3}m}) \propto \chi^{e}\textrm{cos}(3\theta)$, where $\theta$ is taken to be with respect to the $y$-axis. The maximum of the magnetic contribution, $M(\underline{\bar{3}m})$, is rotated by 90 degrees with respect to the electric contribution, $P(\underline{\bar{3}m})$, because of the curl operation in the source term. Above $T_{N}$, the SHG data can be described by just magnetic contribution, while below $T_N$, both electric and magnetic contributions are required to fit the data.  $\chi^{m}$ and $\chi^{e}$ are complex quantities, but only the relative phase between them is important. The relative phase can be obtained by two methods. First, it can be obtained directly from the fit of the rotational anisotropy second harmonic generation (RA-SHG) patterns. The electric and magnetic contribution have roughly an 85 degree phase difference at 1180 nm (the wavelength used in the main manuscript) due to local field effects.  The fact that the nodes are lifted below $T_N$ is due to this relative phase between the two contributions. Second, the phase can be computed using intensity of the left and right circularly polarized second harmonic emission in combination with intensity of the linearly polarized second harmonic emission as shown in the following section.

\subsection{Left and right circular probe}

In this section, we show the two AFM domains can be distinguished at equilibrium by using left and right circularly polarized incident light. Compared to rotational anisotropy SHG measurements, here we replace the polarizer for the incident light with a quarter wave plate that can generate circularly polarized light. We also remove the analyzer for the second harmonic light. The source term in linear basis is given by

\begin{equation*}
\boldsymbol{S} \propto \left(\nabla \times \frac{\partial \boldsymbol{M}}{\partial t}+\frac{\partial^{2} \boldsymbol{P}}{\partial t^{2}}\right)\propto 
\begin{pmatrix}
\chi^{m}(E_{x}^{2}-E_{y}^{2})+2\chi^{e}E_{x}E_{y}  \\
\chi^{e}(E_{x}^{2}-E_{y}^{2})-2\chi^{m}E_{x}E_{y}  \\
0 
\end{pmatrix}
\end{equation*}
We can convert the source term into circular basis with $\boldsymbol{E}=E_{R}\textbf{e}_{R}+E_{L}\textbf{e}_{L}+E_{z}\textbf{e}_{z}$, where $
 \textbf{e}_{R}=-1/\sqrt{2}(\textbf{e}_{x}+i \textbf{e}_{y})$ and $\textbf{e}_{L}=1/\sqrt{2}(\textbf{e}_{x}-i \textbf{e}_{y})$. Then one gets 
 
 \begin{equation*}
\boldsymbol{S}=\begin{pmatrix}
S_{R}\\
S_{L}\\
S_{z}
\end{pmatrix}
\propto  
\begin{pmatrix}
(-\chi^{m}+i\chi^{e}) E_{L}^{2} \\
(\chi^{m}+i\chi^{e}) E_{R}^{2}  \\
0 
\end{pmatrix}
\end{equation*}
 $ I_{R/L} \propto  |\boldsymbol{S}|^{2}$. Hence, the SHG intensity is 
 \begin{equation}
    I_{R/L} \propto  \left(|\chi^{m}|^2+|\chi^{e}|^2\right)\mp 2\left(|\chi^{m}| |\chi^{e}|\right)\textrm{sin}(\gamma ), 
    \label{eq:S1}
   \end{equation} where $\gamma$ is the phase difference between $\chi^{m}$ and $\chi^{e}$. The first term is quadratic in $\chi^{m}$ and $\chi^{e}$ and is
always positive, while the second term is the so-called interference term and may be positive or negative. The sign of the interference term can be switched either by changing the handedness of the circular light or by probing the opposite antiferromagnetic domain. Additionally, the interference is also affected by the wavelength of the probe. The largest contrast in the second harmonic response to the incident left and right circularly polarized light occurs when the magnitude of $\chi^{m}$ and $\chi^{e}$ are the same and possess a 90 degree phase difference.

\subsection{Computing the relative phase}
The relative phase between  $\chi^{m}$ and $\chi^{e}$ can be obtained either from directly fitting the SHG pattern or  derived from Eq.~\ref{eq:S1} as \begin{equation}
    \gamma = \textrm{sin}^{-1}\left(\frac{I_R - I_L}{I_R+I_L}\cdot\frac{I_{M}+I_{E}}{2\sqrt{I_{M}I_{E}}}\right).
    \label{eq:phase}
\end{equation}
Here, the $I_R$ and $I_L$ denote the SHG intensities due to right and left circularly polarized light, and $I_{M}$ and $I_{E}$ denote the SHG intensities due to the magnetic dipole and electric dipole radiation (i.e.$\propto |\chi^{m}|^2$ and $|\chi^{e}|^2$), respectively. The phase obtained using both this method and fitting the RA-SHG patterns are consistent.

\section{analysis of non-equilibrium SHG }
\subsection{Derivation of the $\mathbf{-2\phi}$ in optical rectification process}

The optical rectification caused by the pump is given by $\boldsymbol{P}(\omega_{0})=\boldsymbol{\chi}^{e}_{OR}\boldsymbol{E}^{p}(\omega)\boldsymbol{E}^{p}(-\omega)$.  Under the magnetic point group symmetry $\underline{\bar{3}m}$, the relevant susceptibility tensor elements are $\chi_{yyy}$=$-\chi_{yxx}$=$-\chi_{xyx}$=$-\chi_{xxy}$=$\chi^{e}_{OR}$.
The light induced polarization can be decomposed into $x$ and $y$ components $\boldsymbol{P}(\omega_{0})=P_{x} \textbf{e}_{x}+P_{y} \textbf{e}_{y}$ with $P_{x}=2\chi_{xyx}E_{y}E_{x}$ and $P_{y}=\chi_{yyy}E_{y}E_{y}+\chi_{yxx}E_{x}E_{x}$. For a linearly polarized pump we have $E_{x}=E^{p} \textrm{sin}\phi$ and $E_{y}=E^{p}\textrm{cos}\phi$, where the $E^{p}$ is the magnitude of the electric field from the pump and the $\phi$ is the angle between the pump polarization and the $y$-axis, as defined in the main text.  Therefore, we obtain
\begin{equation}
 \boldsymbol{P}(\omega_{0})=\chi^{e}_{OR}  (E^{p})^2 [\textrm{sin}(-2\phi)\textbf{e}_{x}+ \textrm{cos}(-2\phi)\textbf{e}_{y}]
\label{eq:2}
\end{equation}

\subsection{Functional form of pump induced electric and magnetic dipole contributions to the second harmonic intensity}
The pump induced nonlinear magnetization and polarization are given by 
\begin{equation} 
\begin{split}
 \boldsymbol{M}^{p}(2\omega)= &  \boldsymbol{\chi}^{m} \boldsymbol{E}(\omega) \boldsymbol{E}(\omega) \boldsymbol{P}(\omega_{0})\\
 \boldsymbol{P}^{p}(2\omega)= &  \boldsymbol{\chi}^{e} \boldsymbol{E}(\omega) \boldsymbol{E}(\omega) \boldsymbol{P}(\omega_{0}),
\end{split}
\end{equation} where the $\boldsymbol{\chi}^{e/m}$ are third order susceptibility tensors constrained by the $\underline{\bar{3}m}$ symmetry. The pump gives rise to additional contributions to the second order susceptibility tensors $\delta \chi_{ijk}^{m}=\chi_{ijkl}^{m}P_{l}(\omega_{0})$ and $\delta \chi_{ijk}^{e}=\chi_{ijkl}^{e}P_{l}(\omega_{0})$ that modify the equilibrium tensors, $\chi_{ijk}^{m}$ and $\chi_{ijk}^{e}$. When the pump and probe are normally incident on the sample, only eight matrix elements of the second order tensor, $\chi_{ijk}$, are experimentally accessible. The tensor elements related to the $z$ components can be ignored. We can then write the effective equilibrium second order susceptibility tensor as:

\begin{equation*}
      \boldsymbol{\chi}^{e/m} = \begin{pmatrix} 0 & -\chi^{e/m}  \\
            -\chi^{e/m} & 0    \\
            -\chi^{e/m} & 0 \\
            0 & \chi^{e/m}\\
            \end{pmatrix}  \\
\end{equation*}
When considering the effect of the pump, we therefore need to consider only eight tensor elements of $\chi^{e/m}_{ijkl}$:

\begin{equation*}
\left\lbrace \begin{array}{l}
\chi_{xxyy}=\chi_{yyxx}\\
\chi_{xyyx}=\chi_{yxxy}\\
\chi_{xyxy}=\chi_{yxyx}\\
\chi_{xxxx}=\chi_{yyyy}=\chi_{xxyy}+\chi_{xyyx}+\chi_{xyxy}
\end{array}\right.\end{equation*}
The perturbation due to induced polarization can then be included into an effective second order susceptibility tensor by writing:

\begin{align*}
       \delta\boldsymbol{\chi}^{e/m} = \begin{pmatrix} \chi_{xxxx}\cdot P_{x} & \chi_{xyxy}\cdot P_{y}  \\
            \chi_{xxyy}\cdot P_{y} & \chi_{xyyx}\cdot P_{x}    \\
             \chi_{yxxy}\cdot P_{y} & \chi_{yyxx}\cdot P_{x} \\
       \chi_{yxyx}\cdot P_{x} & \chi_{yyyy}\cdot P_{y}  \\
       \end{pmatrix}\\
\end{align*} 
From the previous section, we have  $\boldsymbol{P}(\omega_{0})=P_{x} \textbf{e}_{x}+P_{y} \textbf{e}_{y}=\chi^{e}_{OR}  (E^{p})^2 [\textrm{sin}(-2\phi)\textbf{e}_{x}+ \textrm{cos}(-2\phi)\textbf{e}_{y}]$. Then we obtain the  $M(C_{2}) \propto \delta\chi^{m}\textrm{sin}(\theta-2\phi)$ and $P(C_{2}) \propto \delta\chi^{e}\textrm{cos}(\theta-2\phi)$ as mentioned in the main text, where $\delta\chi^{m} \propto \chi^{m}_{yyyy}\chi^{e}_{OR}(E^{p})^{2}$ and $\delta\chi^{e} \propto \chi^{e}_{yyyy}\chi^{e}_{OR}(E^{p})^{2}$.
In summary, we have 
\begin{equation}
\begin{split}
I(2\omega; &\varphi) \propto |e^{i\gamma}\chi^m\textrm{sin}(3\theta) + \chi^e\textrm{cos}(3\theta) \\
&+ie^{i\gamma}\delta\chi^m\textrm{sin}(\theta-2\varphi) + i\delta\chi^e\textrm{cos}(\theta-2\varphi)|^2.
\label{eq:PerturbationPhase}
\end{split}
\end{equation}

\subsection{Property of susceptibility tensors and complexity of each contribution}

For physical fields, $A(t)$, we have $\Theta A(t)=\pm A(t)$, where $\Theta$ represents the time reversal operator.  The ``$+$'' sign corresponds to i-type fields and ``$-$'' sign to c-type fields. In the Fourier domain, we get the corresponding relationship: $\Theta A(\omega)=\pm A^{*}(\omega)$. In Cr$_{2}$O$_{3}$, below $T_N$, inversion and time reversal symmetry are both broken by the antiferromagnetic
arrangement of spins, but the combined symmetry of inversion and time reversal remains
a symmetry element \textit{in the absence of dissipation}. Hence we have 
\begin{equation}
 \Theta \mathbf{I} \boldsymbol{\chi}^{e/m}=\boldsymbol{\chi}^{e/m},
\label{eq:TI}
\end{equation} where $\Theta$ and $\mathbf{I}$ are the time reversal and inversion operators, respectively. Applying the combined time reversal and inversion operators  $\Theta \mathbf{I}$ to $\boldsymbol{P}(\omega_0)=\boldsymbol{\chi}^{e}_{OR}\boldsymbol{E}(\omega_1)\boldsymbol{E}(-\omega_2)$, we get  $-\boldsymbol{P}^{*}(\omega_0)=\boldsymbol{\chi}^{e}_{OR}(-\boldsymbol{E}^{*}(\omega))(-\boldsymbol{E}^{*}(\omega))$, where we have used Eq.~\ref{eq:TI}. After taking the complex conjugate of this equation, we get  $\boldsymbol{P}(\omega_0)=-(\boldsymbol{\chi}^{e}_{OR})^{*}\boldsymbol{E}(\omega_1)\boldsymbol{E}(-\omega_2)$. The susceptibility tensor describing optical rectification is then constrained to obey $\boldsymbol{\chi}^{e}_{OR}=-(\boldsymbol{\chi}^{e}_{OR})^{*}$, indicating $\boldsymbol{\chi}^{e}_{OR}$ is purely imaginary. Therefore, in the absence of dissipation (as satisfied by pumping below the gap), the light-induced polarization $\boldsymbol{P}(\omega_0)$ is purely imaginary, and $\delta\chi^{e}$ ($\delta\chi^{m}$) and $\chi^{e}$ ($\chi^{m}$) must possess a 90 degree phase difference. This is consistent with our results and can explain the domain contrast observed upon pumping.

Above the Néel temperature $T_{N}$, Cr$_{2}$O$_{3}$ possesses symmetries of the centrosymmetric point group $\bar{3}m$. Axial i-type tensors of odd rank, such as $\chi_{ijk}^{m}$, and polar i-type tensors of even rank, such as $\chi_{ijkl}^{e}$, are allowed. Below $T_{N}$, inversion symmetry is broken due to the spin arrangement and the magnetic point group of Cr$_{2}$O$_{3}$ is $\underline{\bar{3}m}$. In addition to $\chi_{ijk}^{m}$ and $\chi_{ijkl}^{e}$, 
polar c-type tensors of odd rank, such as $\chi_{ijk}^{e}$,  and axial c-type tensors of even rank, such as $\chi_{ijkl}^{m}$, become allowed. Both of these tensors are proportional to the order parameter (i.e. the Néel vector) and flip sign when examining opposite AFM domains. 
Therefore, $\delta \chi^{e}$ is an axial c-type tensor and $\delta \chi^{m}$ is a polar i-type tensor. Hence, $\delta \chi^{e}$ flips sign when examining opposite antiferromagnetic domains, while $\delta \chi^{m}$ remains unchanged.

\section{Temperature dependence of SHG}
\begin{figure}[hbt!]
	\includegraphics[width=0.9\columnwidth]{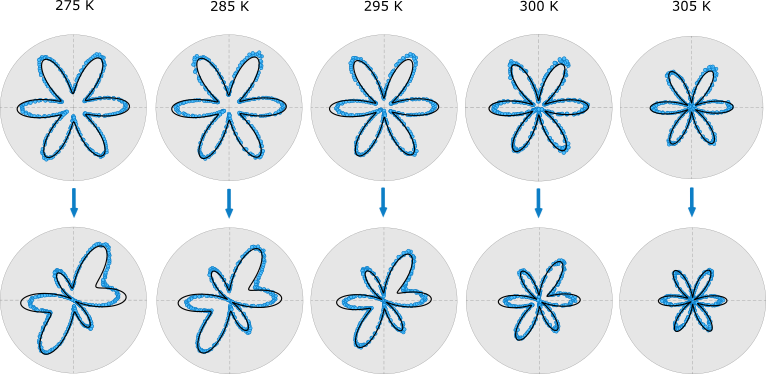}
	\caption{Top row: unpumped SHG pattern of $\beta$ domain at various temperatures. Bottom row: pumped SHG pattern taken at time zero at various temperatures. }
	\label{FigS1}
\end{figure}

\begin{figure}[hbt!]
	\includegraphics[width=0.9\columnwidth]{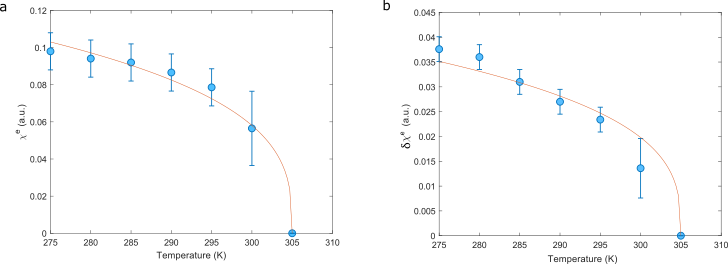}
	\caption{ (a)  $\chi^{e}$  versus temperature. The solid curve is the fit using $a(T-T_{c})^{\beta}$, the obtained $\beta \sim$ 0.32 is consistent with previous results, confirming  the c type tensor $\chi_{ijk}^{e}$ is proportional to order parameter. (b)  $\delta\chi^{e}$  versus temperatures. The solid line is fitting by fixing  $\beta =$ 0.32.}
	\label{FigS2}
\end{figure}

In this section, we present pumped and unpumped RA-SHG patterns at various temperatures, and show that the pump induced susceptibility tensor $\delta\chi^{e}$ has an order parameter-like behavior as a function of temperature. We plot the unpumped and pumped SHG patterns in Fig.~\ref{FigS1} at five selected temperatures close to $T_{N} \sim 307$~K. Laser heating increases the sample temperature by roughly 3-5~K. Therefore, for the 305~K data, the actual temperature is above  $T_{N}$ as evidenced by the appearance of the nodes in RA-SHG pattern (and equal SHG intensities when examining the left and right circularly polarized SHG signals). From our fits, we can extract $\chi^{e}$ from the unpumped data and $\delta\chi^{e}$ from the pumped data at each temperature. In Fig.~\ref{FigS2}, we present $\chi^{e}$ and $\delta\chi^{e}$ versus temperature. We find that, consistent with previous studies, $\chi^{e}$ is proportional to the Néel vector. The critical exponent of the order parameter $\beta\approx$ 0.32. In the previous section we showed that $\delta\chi^{e} \propto \chi^{e}_{yyyy}\chi^{e}_{OR}$, which indicates that  $\delta\chi^{e}$ is also related to the order parameter and should also disappear above  $T_{N}$. Under the assumption that the i-type tensor element $\chi^{e}_{yyyy}$ is constant across the AFM transition, $\delta\chi^{e}$ is linear proportional to the order parameter. It should be kept in mind, however, that $\chi^{e}_{yyyy}$ may scale quadratically with the order parameter. From our data alone, it is difficult to ascertain whether $\delta\chi^{e}$ is linearly proportional to the order parameter due to the large measurement uncertainty. The best fit gives rise to an exponent $\beta \sim$ 0.5, but $\beta \sim$0.32 also falls within the error bars. Nevertheless, the pump induced  $\delta\chi^{e}$ is related to the order parameter and should flip signs as switching to the other domain.

\newpage
\vfill

\section{SHG in the perpendicular channel}
\begin{figure}[hbt!]
	\includegraphics[width=0.8\columnwidth]{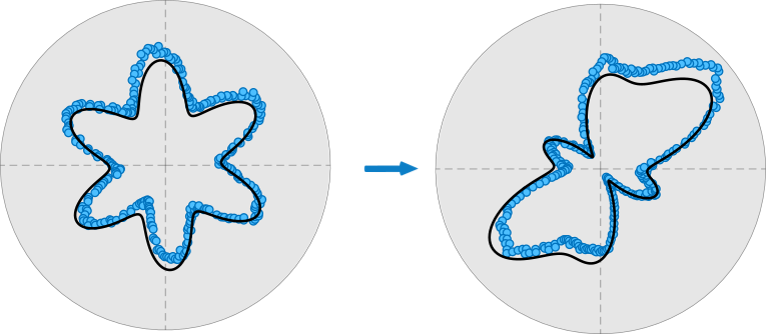}
	\caption{The equilibrium and non-equilibrium SHG pattern of  $\beta$  domain for perpendicular channel. The probe wavelength 1180 nm and the pump is along x axis.  }
	\label{FigS4}
\end{figure}

\begin{figure}[hbt!]
	\includegraphics[width=0.9\columnwidth]{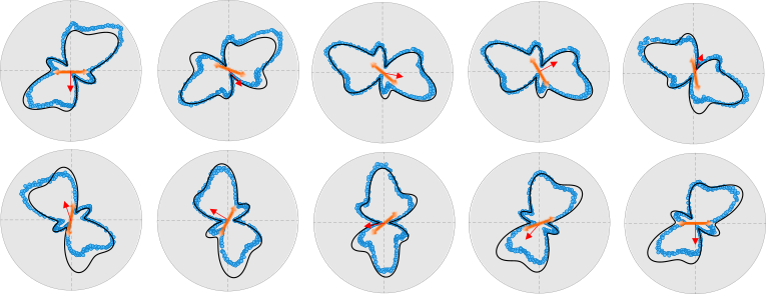}
	\caption{ The evolution of SHG pattern  at time zero for  $\beta$  domain as varying the pump polarization. The pump polarization starting along x axis is rotated clockwise at 20$^{\circ}$  step size, denoted by the orange double headed arrow. The C$_{2}$ symmetry axis is represented by red single headed arrow. Note only the solid black for pump along x is fit,  the rest is obtained by varying $2\phi$ in opposite direction and keeping all the other parameters the same.}
	\label{FigS5}
\end{figure}

In the main text and previous section, we showed data with the two polarizers (for the incident and outgoing light) in a parallel geometry. Here, we show the data in the perpendicular geometry. In our setup, we rotate the half wave plate so that the input polarization is rotated by 90 degrees to achieve the perpendicular geometry. In contrast to the parallel geometry, we now get $M(\underline{\bar{3}m}) \propto \chi^{m}\textrm{cos}(3\theta)$ and $P(\underline{\bar{3}m}) \propto \chi^{e}\textrm{sin}(3\theta)$ for the equilibrium RA-SHG pattern, while the terms $M(C_{2}) \propto \delta\chi^{m}\textrm{cos}(\theta-2\phi)$ and $P(C_{2}) \propto \delta\chi^{e}\textrm{sin}(\theta-2\phi)$ need to be included upon pumping. In the perpendicular channel, $\delta\chi^{m} \propto \chi^{m}_{yxxy}\chi^{e}_{OR}(E^{p})^{2}$ and $\delta\chi^{e} \propto \chi^{e}_{yxxy}\chi^{e}_{OR}(E^{p})^{2}$. Fig.~\ref{FigS4} shows the unpumped RA-SHG pattern with a fundamental probe wavelength of 1180~nm in the perpendicular channel. This figure also shows the corresponding pumped RA-SHG pattern with pump polarization along the $x$ axis. Fig.~\ref{FigS5} shows the pumped RA-SHG pattern as the pump polarization is rotated clockwise in 20 degree steps. Again, these are not fits to the individual RA-SHG patterns -- instead, only the $\varphi = -90^\circ$ pattern is fit and the black curves pertaining to the remaining RA-SHG patterns are generated from our fit function. 

\newpage
\section{Effect of circular pump}
In this section, we demonstrate that a circularly polarized pump pulse does not induce a quasi-static electric dipole in Cr$_2$O$_3$ when pumping in the transparency window away from any electronic transitions. 
This calculation is more easily demonstrated in the time-domain, where we make the assumption that the medium is lossless and dispersionless. We can write that $\boldsymbol{\tilde{P}}(t)=\boldsymbol{\chi}^{e}_{OR}\boldsymbol{\tilde{E}}^{p}(t)\boldsymbol{\tilde{E}}^{p}(t)$, where the tilde denotes a time-dependent quantity. Under the magnetic point group symmetry $\underline{\bar{3}m}$, the relevant susceptibility tensor elements are   $\chi_{yyy}$=$-\chi_{yxx}$=$-\chi_{xyx}$=$-\chi_{xxy}$=$\chi^{e/m}$. 
We can write $\boldsymbol{\tilde{P}}(t)=\tilde{P}_{x} \textbf{e}_{x}+\tilde{P}_{y} \textbf{e}_{y}$, and the $x$ and $y$ components are $\tilde{P}_{x}=2\chi_{xyx}\tilde{E}_{y}\tilde{E}_{x}$ and $\tilde{P}_{y}=\chi_{yyy}\tilde{E}_{y}\tilde{E}_{y}+\chi_{yxx}\tilde{E}_{x}\tilde{E}_{x}$ as in the previous section. Now with a circularly polarized pump, the $x$ and $y$ components of pump electric field are out of phase, and we have (for right circular pump) $\tilde{E}_{x}=\frac{1}{\sqrt{2} }(E^{p})\textrm{sin}(\omega t) $ and $\tilde{E}_{y}=\frac{1}{\sqrt{2} } (E^{p}) \textrm{cos}(\omega t)$ , where the $E^{p}$ is the electric field from the pump.   We obtain
\begin{equation}
 \boldsymbol{\tilde{P}}(t)=\frac{1}{2}\chi^{e}_{OR}  (E^{p})^2 [\textrm{sin}(-2\omega t)\textbf{e}_{x}+\textrm{cos}(-2\omega t)\textbf{e}_{y}].
\label{eq:S7}
\end{equation}
Because the calculation does not consist of a constant part (i.e. a time-independent contribution), but only a time-dependence at twice the incident frequency, there is no rectified signal. Circularly polarized light only generates second harmonic light, and not optical rectification, in Cr$_2$O$_3$.

\begin{figure}[hbt!]
	\includegraphics[width=0.7\columnwidth]{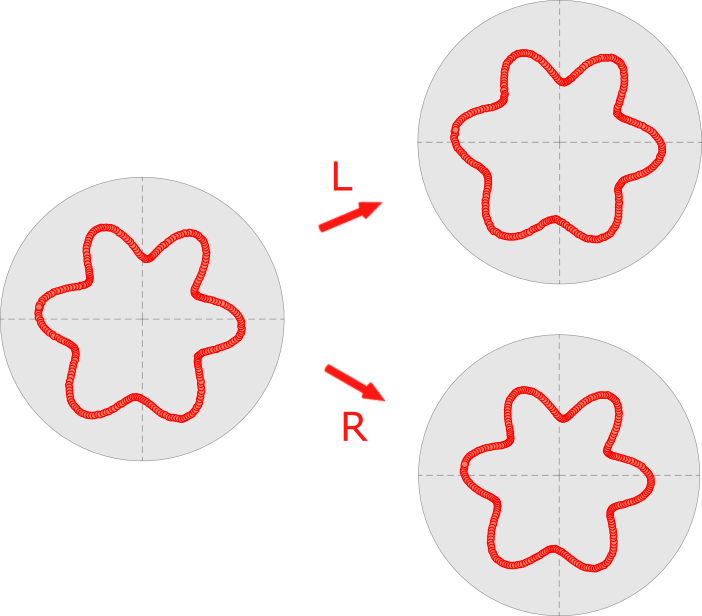}
	\caption{Both left and right circular pump has minimal change to the SHG pattern of $\alpha$ domain. The probe wavelength is 1180 nm. }
	\label{FigS6}
\end{figure}

In Fig.\ref{FigS6}, we show that when the $\alpha$ domain is pumped with circularly polarized light, there is no change in symmetry, consistent with the calculation presented above. The circularly polarized light was generated using quarter wave plate and the fluence used was approximately $20$~mJ/cm$^{2}$, the same as that used in the main manuscript with linearly polarized light.

\newpage
\section{wavelength dependence and phase}
\begin{figure*}[t]
	\centering
	\includegraphics[width=0.85\columnwidth]{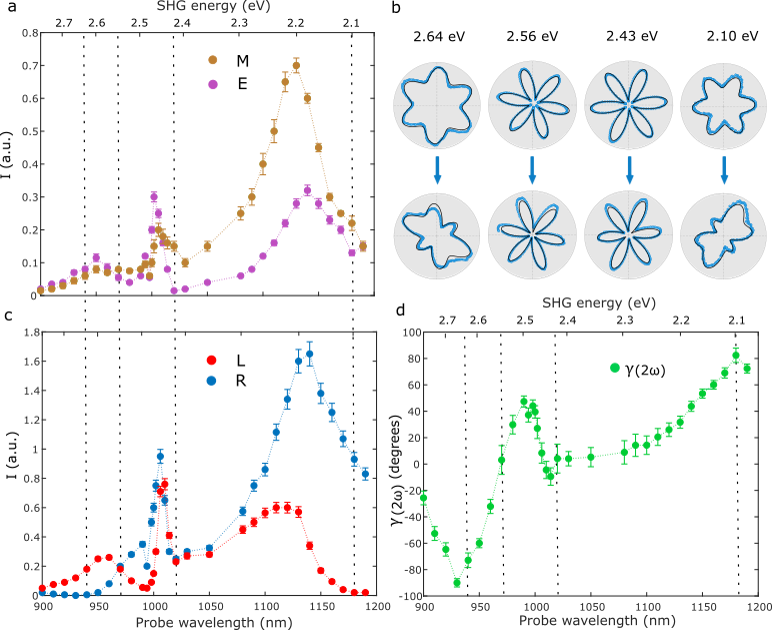}
	\caption{(a) Wavelength dependence of SHG intensity for magnetic $I_{M} \propto |M(\underline{\bar{3}m})|^{2}$ and electric contribution  and $I_{P} \propto |P(\underline{\bar{3}m})|^{2}$ at equilibrium, measured with probe polarization  set to be along x and y axis, respectively.   (b) Unpumped and pumped SHG patterns at four representative wavelengths 940, 970, 1020, 1180 nm, corresponding to SHG energy 2.6, 2.55, 2.4, 2.1 eV. The probe intensity is not calibrated for each SHG pattern.(c) Wavelength dependence of SHG signal probed with left and right circular light at equilibrium. (d) The computed relative phase difference for magnetic and electric radiation. The wavelength dependence measurements were done on  $\beta$  domain. }  
	\label{fig:S7}
\end{figure*} 
In this section, we highlight the key role played by the equilibrium interference between electric and magnetic dipole radiation in observing the symmetry-breaking effects in Cr$_2$O$_3$. As we tune the second harmonic energy across several electronic $d$-$d$ transitions, the relative phase difference  $\gamma$ between electric and magnetic radiation changes correspondingly, and can be experimentally determined using Eq.~\ref{eq:phase} and is shown in Fig.~\ref{fig:S7}(d).

In the top panel of Fig.~\ref{fig:S7}(b), we show four equilibrium RA-SHG patterns at representative second harmonic energies corresponding to the dashed vertical lines in Fig.\ref{fig:S7}(a), (c) and (d). These values were chosen because they constitute extremal values of $\gamma$ ($\sim$0$^\circ$ at 2.56~eV and 2.43~eV, $\sim90^\circ$ at 2.64~eV and $\sim-90^\circ$ at 2.10~eV). In Fig.~\ref{fig:S7}(b), we show that when $\gamma \approx 0^\circ$, nodes are observed in the equilibrium RA-SHG pattern, and the induced dipole is barely visible when pumped. On the other hand, when $\gamma \approx \pm90^\circ$, the equilibrium nodes are lifted, and the pump has a dramatic effect on the symmetry of the RA-SHG. That the effect is most drastic at $\gamma\approx\pm90^\circ$ is also captured by Eq.~\ref{eq:PerturbationPhase}.  When $\gamma\neq0^\circ$, interference between the equilibrium and perturbing terms arise. These interference terms are linear in $\delta \chi^{e/m}$. In contrast, when $\gamma=0^\circ$ Eq.~\ref{eq:PerturbationPhase} factorizes and only terms $\mathcal{O}(\delta^2)$ remain. The lack of interference thus renders the induced dipole imperceptible.
\newpage

\section{time dependent RA-shg }

\begin{figure*}[t]
	\centering
	\includegraphics[width=0.9\columnwidth]{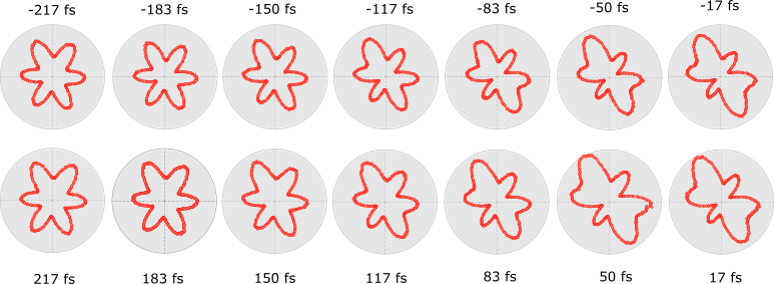}
	\caption{Representative rotational anisotropy patterns of the emitted second harmonic light as a function of time in the $\alpha$ domain when pumping along $x$ axis. }  
	\label{fig:S8}
\end{figure*}

In Fig.~\ref{fig:S8}, we show how the rotational anisotropy pattern of the second harmonic light evolves as a function of time as the pump pulse propagates through Cr$_2$O$_3$. Specifically, we show the evolution of the RA-SHG in the $\alpha$ domain. The change due to the pump pulse (at a fluence of $\sim$20~mJ/cm$^2$) is roughly symmetric about $t=0$. This symmetry about $t=0$, which would not be anticipated if the spin or lattice degrees of freedom were involved, is evidence of a purely electronic process that breaks the underlying symmetry of the lattice.

\newpage
\section{wave-functions for microscopic mechanism of optical rectification }
At equilibrium,  the ground state $| g\rangle$ in  Cr$_2$O$_3$ is $^{4}\hspace{-0.1cm}A_{2g}$, which comprises three electrons in $t_{2g}$ state.   The relevant excited states, $| e\rangle$, comprise the $^{4}T_{2g}$ and $^{4}T_{1g}$ levels, both of which possess two $t_{2g}$ electrons and one $e_{g}$ electron. The $t_{2g}$ and  $e_{g}$ states can be constructed based on the $\underline{\bar{3}m}$ point group symmetry as shown by Eq.~\ref{eq:wavefunction}:

\begin{equation}
\begin{split}
 &\ket{t_{2g}^{(1)}}=d_{x^2+y^2}\\
 & \ket{t_{2g}^{(2)}}=((id_{xy}-d_{x^2-y^2}+d_{zx}+id_{yz})+ \eta(p_{x}+i p_{y}))/(1+\eta)\\
 &\ket{t_{2g}^{(3)}}=((id_{xy}+d_{x^2-y^2}-d_{zx}+id_{yz})+ \eta(-p_{x}+i p_{y}))/(1+\eta)\\
&\ket{e_{g}^{(2)}}=((id_{xy}+d_{x^2-y^2}+d_{zx}-id_{yz})+ \eta(p_{x}-i p_{y}))/(1+\eta)\\
&\ket{e_{g}^{(1)}}=((id_{xy}-d_{x^2-y^2}-d_{zx}-id_{yz})+\eta (-p_{x}-i p_{y}))/(1+\eta)
\end{split}
\label{eq:wavefunction}
\end{equation}

where, for this choice of basis, $z$ is along the direction of the trigonal elongation, (i.e. the [111]-direction of the oxygen octahedron). Here, we already take into consideration the mixing of 4p orbitals due to the trigonal field, where $\eta$ denotes the strength of trigonal distortion and an arbitrary number is used for illustration purposes ($\eta =0.7$). As will be shown in next section, a single-electron approach is equivalent to multi-electron approach. Therefore the total electric dipole can be obtained by adding up the dipole moments of each  single electron.  We can draw  the corresponding single electron probability densities (in the $x-y$ plane) for $| g\rangle$ and $| e\rangle$ based on the above wave-functions. In the Fig.~4(a), we show the electron probability densities of ($t_{2g}^{(1)}$, $t_{2g}^{(2)}$, $t_{2g}^{(3)}$) for $| g\rangle$ and one combination $(t_{2g}^{(1)}, t_{2g}^{(2)}, e_{g}^{(1)})$ as an example for the $| e\rangle$ manifold. The $d_{zx}$ and $d_{yz}$ do not contribute to the probability densities in the $x-y$ plane.

In the presence of time periodic pump pulse, Floquet states are created and we show only the lowest order (within the rotating wave approximation) $| e,-1\rangle$ state in Fig.~4. There are two pathways that can give rise to a static dipole as explained in the main text. 
The consequences of these two pathways are the mixing among $(t_{2g}^{(1)}, t_{2g}^{(2)}, t_{2g}^{(3)}, e_{g}^{(1)}, e_{g}^{(2)})$ states in terms of the single electron approach. We show a representative electron probability density for the hybridized state $\ket{g,0} + \epsilon^2\ket{e,0}$ (corresponding to the first pathway): $|t_{2g}^{(1)}\rangle \rightarrow|t_{2g}^{(1)}+ \zeta(t_{2g}^{(2)}+t_{2g}^{(3)}) + \xi e_{g}^{(1)}\rangle $, $|t_{2g}^{(2)}\rangle \rightarrow|t_{2g}^{(2)}+ \zeta (t_{2g}^{(1)}+t_{2g}^{(3)}) + \xi e_{g}^{(1)}\rangle $ and $|t_{2g}^{(3)}\rangle \rightarrow|t_{2g}^{(3)}+ \zeta (t_{2g}^{(1)}+t_{2g}^{(2)}) + \xi e_{g}^{(1)}\rangle $. 
The coefficients characterizing the hybridization strength are determined by transition matrix elements (such as intensity and light polarization). Arbitrary values $\xi$ and $\zeta$  are chosen for illustration purposes, as shown in Fig.4 in the main text (e.g.,  $\zeta =0.1$ and $\xi =0.2$ ). As shown in Fig.4, the hybridized states break the threefold in-plane rotational symmetry and can give rise to the light-induced electronic dipole moment.

\section{Perturbative calculation of dipole moment}

We set up a toy model for the pumped Cr$_2$O$_3$ and calculate the static dipole moment using time-dependent perturbation theory.  Then, we repeat the same calculation using the Floquet approach.  Finally, we show that the same answer must be obtained if the dipole moment is written as the sum of dipole moments in time-evolving single-electron states.

\subsubsection{Time-dependent perturbation theory}
We consider an atomic model consisting of a non-degenerate ground state $\ket{0}$ at energy $w_0$ (we set $\hbar=1$ throughout) and excited states $\ket{1},\dots,\ket{N}$ at energy $w_1=\dots=w_N  > w_0$, with dipole coupling to light.  Thus, we consider
\beq
    H(t) = H^{(0)}+ H^{(1)}(t),
\eeq
where $H^{(0)}$ is the unperturbed Hamiltonian and $H^{(1)}(t)$ is the dipole coupling to the pump field $\mathbf{E}(t)$: 

\bseq
\begin{align}
    H^{(0)}&= \sum_{i=0}^N w_i \ket{i}\bra{i}\\
    H^{(1)}(t) &= - \mathbf{d}\cdot \mathbf{E}(t).
\end{align}
\eseq
We do not include dissipation because the experiment was done in a regime without significant absorption. 

Setting $N=6$ corresponds to a toy model of a given Cr atom in Cr$_2$O$_3$.  The ground state is the $^{4}\hspace{-0.1cm}A_{2g}$ state ($\ket{g}$ in the main text), and the only excited states kept are the three $^{4}T_{1g}$ states and the three $^{4}T_{2g}$ states ($\ket{e}$ in the main text).  We need not consider excited states that flip the spin because they are dipole forbidden.  In this toy model, we assume that the six excited states are exactly degenerate.  While the energy splitting (approximately $0.6$ eV) between the $^{4}T_{1g}$ states and $^{4}T_{2g}$ states may have a non-negligible quantitative effect on the static dipole moment, we set this splitting to zero as a simplifying assumption to convey the basic idea.  Non-zero splitting can be included straightforwardly in this approach.

A perturbative treatment of $H^{(1)}$ is appropriate because the linear dependence on fluence [Fig. 4(b) in the main text] indicates that the static dipole depends quadratically on the electric field amplitude.  Also, we can roughly estimate the electric field amplitude $E$ from the fluence $f=20$ mJ/cm$^2$ via $f\approx |\frac{1}{\mu_0}\mathbf{E}\times\mathbf{B}| t_\text{pulse} = c\epsilon_0 E^2 t_\text{pulse}$  (where $t_\text{pulse}=180$ fs), yielding $E\approx 6 * 10^8$ V/m. Then, estimating the dipole moment as the electron charge times the Bohr radius, we obtain the order of magnitude of $H^{(1)}(t)$ as $0.06$ eV, which is small compared to $w_2-w_1=2.1$ eV (the energy gap between $^{4}T_{2g}$ and $^{4}\hspace{-0.1cm}A_{2g}$).

We take the electric field $\mathbf{E}(t)$ of the pump pulse to be linearly polarized:
\beq
    \mathbf{E}(t) = \varphi(t)\mathbf{E} =  \int\frac{dw}{2\pi}\ e^{-i wt}\varphi(w)\mathbf{E},
\eeq
where $\mathbf{E}$ is a fixed vector defining the polarization axis, $\varphi(t)$ is the time dependence of pump pulse, and $\varphi(w) = \int dt\ e^{i wt} \varphi(t)$ is the pump pulse in the frequency domain.

Since $\varphi(t)$ must be real, $\varphi(w) = \varphi^*(-w)$; we therefore focus on $w>0$ in the following discussion of the pump pulse.  We take $\varphi(w)$ to be sharply peaked, with some pulse width $\Delta$, about the laser frequency $\Omega$; in particular, we assume
\beq
    \varphi(w) \overset{\Delta\to0^+}{\longrightarrow} \pi \delta(w-\Omega) \qquad (w>0).
\eeq
Thus, $\varphi(t)\overset{\Delta\to0^+}{\longrightarrow} \cos(\Omega t)$, while for small $\Delta>0$, $\varphi(t)$ is a broad wave train at frequency $\Omega$ that goes to zero at $t\to\pm\infty$.  We are interested in time $t$ in the broad middle of the wave train.  For simplicity, we assume that $\varphi(w)$ is strictly zero for $w$ outside of $[\Omega-\frac{1}{2}\Delta,\Omega+\frac{1}{2}\Delta]$.

We consider evolving the system starting from the ground state at time $t_0$ (later sending $t_0\to-\infty$).  We wish to calculate the dipole moment at a time $t$:
\beq
    \langle \mathbf{d}\rangle_t \equiv \bra{\Psi(t)}\mathbf{d} \ket{\Psi(t)},
\eeq
where
\beq
    \ket{\Psi(t)} = U(t,t_0)\ket{0}= \sum_{i=0}^N a_i(t) \ket{i}.
\eeq

We calculate the coefficients $a_i(t)$ perturbatively:
\beq
    a_i(t) = a_i^{(0)}(t)+a_i^{(1)}(t) + a_i^{(2)}(t)+ \dots,
\eeq
where the superscripts indicate the order in $|\mathbf{E}|$.  It is convenient to first calculate the coefficients in the interaction picture (relative to $t=t_0$), i.e., to define
\beq
    a_{i,I}(t) = e^{i w_i (t-t_0)}a_i(t).
\eeq
The interaction picture coefficients are determined order-by-order by $a_{i,I}^{(n)}(t_0) = \delta_{i0}\delta_{n 0}$ and

\bseq
\begin{align}
    i\frac{d}{d t}a_{i,I}^{(n+1)}(t)&= \sum_{j=0}^N [H_I^{(1)}(t)]_{ij} a_{j,I}^{(n)}(t)\\
    &= -\sum_{j=0}^N \mathbf{d}_{ij}\cdot \mathbf{E}\varphi(t) e^{i(w_i - w_j)(t-t_0)} a_{j,I}^{(n)}(t),
\end{align}
\eseq
where we use the notation $\mathcal{O}_{ij}= \bra{i}\mathcal{O}\ket{j}$ for the matrix elements of any operator $\mathcal{O}$.

Throughout this calculation, we always take the vector $\mathbf{d}$ to be in-plane ($x$-$y$).  This is justified because the pump field was polarized in-plane [hence only the in-plane components of $\mathbf{d}$ appear in $H^{(1)}(t)$] and because the measurement could only detect in-plane components of $\mathbf{d}$ [hence we only need to calculate $\langle \mathbf{d}\rangle_t$ in-plane].  Due to the $C_3$ site symmetry of the Cr atom, the in-plane components of the dipole vanish in the ground state:
\beq
    \mathbf{d}_{00}= \mathbf{0}.
\eeq

We work to second order in $|\mathbf{E}|$ because, as we will see, this yields the leading contribution to the static part of the dipole moment.  Once the $a_{i}(t)$ coefficients are determined up to second order, the dipole moment is obtained as
\beq
    \langle \mathbf{d}\rangle_t = \langle \mathbf{d}\rangle_t^{(0)} + \langle \mathbf{d}\rangle_t^{(1)}+\langle \mathbf{d}\rangle_t^{(2)} + \dots,   
\eeq
where one finds, after noting $a_i^{(0)}(t) = \delta_{i0}e^{-iE_0 (t-t_0)}$ and $\mathbf{d}_{00}=\mathbf{0}$ [which imply $a_0^{(1)}(t)=0$],

\bseq
\begin{align}
    \langle \mathbf{d}\rangle_t^{(0)}&= \mathbf{0},\\
    \langle \mathbf{d}\rangle_t^{(1)}&= 2\sum_{i=1}^N\text{Re}\left( e^{iw_0 (t-t_0)}\mathbf{d}_{0 i} a_i^{(1)}(t) \right),\\
    \langle \mathbf{d}\rangle_t^{(2)} &= \sum_{i,j=1}^N  a_i^{(1)*}(t)\mathbf{d}_{i j}a_j^{(1)}(t)+2\sum_{i=1}^N\text{Re}\left( e^{iw_0 (t-t_0)}\mathbf{d}_{0 i} a_i^{(2)}(t)   \right).\label{eq:d2} 
\end{align}
\eseq

We send the start time $t_0\to-\infty$ before sending the pulse width to zero ($\Delta\to 0^+$).  We take these limits with the observation time $t$ held fixed so that $t$ is in the broad middle of the wave train ($\Delta |t| \ll 1$).  We send the pulse width to zero to yield an approximate answer that is independent of the particular way the pulse turns on and off in time.  This results in some contributions to $\langle \mathbf{d}\rangle_t$ being strictly static.  If we instead were to keep a small non-vanishing pulse width, then additional quasistatic contributions would appear (with frequencies no larger than the pulse width) that depend on further details of how the pulse turns on and off in time.

We readily obtain the first-order correction to the wavefunction:

\bseq
\begin{align}
    a_i^{(1)}(t) &\overset{t_0\to-\infty}{\longrightarrow} e^{-iw_0(t-t_0)}  \mathbf{d}_{i0}\cdot\mathbf{E}\int\frac{dw}{2\pi}\ \frac{1}{w_i- w_0- w}e^{-i w t} \varphi(w)\\
    &\overset{\Delta\to 0^+}{\longrightarrow} e^{- iw_0(t-t_0)}\mathbf{d}_{i0}\cdot\mathbf{E}\frac{1}{2}\left( \frac{1}{w_i - w_0 - \Omega} e^{-i\Omega t} + \frac{1}{w_i- w_0 + \Omega} e^{i\Omega t}\right)\\
    &= e^{- iw_0(t-t_0)} \mathbf{d}_{i0}\cdot\mathbf{E}\frac{1}{2}\left( \frac{1}{w_1 - w_0 - \Omega} e^{-i\Omega t} + \frac{1}{w_1- w_0 + \Omega} e^{i\Omega t}\right)
\end{align}
\eseq
where $t_0$ still appears in the overall phase but always cancels in $\langle \mathbf{d}\rangle_t$.  In the last line, we have replaced $w_i\to w_1$ because $\mathbf{d}_{00}=\mathbf{0}$.

The first order part of the dipole moment therefore does not have any static part (the only oscillation frequency is $\Omega$).  So, we proceed to the second order, finding
\beq
    a_i^{(2)}(t)\overset{t_0\to-\infty}{\longrightarrow} e^{-iw_0(t-t_0)}\sum_{j=1}^N (\mathbf{d}_{ij}\cdot\mathbf{E})(\mathbf{d}_{j0}\cdot\mathbf{E})\int \frac{dw}{2\pi}\frac{dw'}{2\pi}\ \varphi(w)\varphi(w')  \frac{1}{w_j - w_0 - w'}\frac{e^{-i(w+w')t}}{w_i-w_0-w-w'}.
\eeq
Except for $i=0$ [which we see from Eq. \eqref{eq:d2} is not needed], we can send the pulse width to zero, yielding:

\bseq
\begin{align}
     a_i^{(2)}(t)&\overset{\Delta\to 0^+}{\longrightarrow}   e^{-iw_0(t-t_0)}\sum_{j=1}^N (\mathbf{d}_{ij}\cdot\mathbf{E})(\mathbf{d}_{j0}\cdot\mathbf{E})\biggr[ \frac{1}{4} \left( \frac{1}{w_j - w_0 - \Omega} +  \frac{1}{w_j - w_0 +  \Omega}\right) \frac{1}{w_i-w_0} + ( e^{\pm i 2  \Omega t} \text{ terms})\biggr]\\
     &= e^{-iw_0(t-t_0)}\sum_{j=1}^N (\mathbf{d}_{ij}\cdot\mathbf{E})(\mathbf{d}_{j0}\cdot\mathbf{E})\biggr[ \frac{1}{2} \frac{1}{w_1 - w_0 - \Omega}  \frac{1}{w_1 - w_0 +  \Omega} + ( e^{\pm i 2  \Omega t} \text{ terms})\biggr].
\end{align}
\eseq
We ignore the $e^{\pm i 2  \Omega t}$ terms because they do not contribute to the static part of the dipole moment at second order.

From Eq. \eqref{eq:d2}, we thus obtain the following expression for the static part of the dipole moment at leading order:

\begin{multline}
     \langle \mathbf{d}\rangle_t \overset{\text{static}}{=} \frac{1}{4}\left[ \left(\frac{1}{w_1-w_0 -\Omega}\right)^2 + \left(\frac{1}{w_1-w_0 +\Omega}\right)^2 \right] \sum_{i,j=1}^N (\mathbf{d}_{0i}\cdot\mathbf{E}) \mathbf{d}_{ij} (\mathbf{d}_{j0}\cdot\mathbf{E})\\ + \frac{1}{w_1-w_0-\Omega}\frac{1}{w_1-w_0+\Omega} \sum_{i,j=1}^N \text{Re}\left[\mathbf{d}_{0i} (\mathbf{d}_{ij}\cdot\mathbf{E})(\mathbf{d}_{j0} \cdot\mathbf{E})\right].\label{eq:static dipole toy model}
\end{multline}

There is a simple interpretation of this expression: the first term is the contribution from the first-order population of the excited states (resulting in a second-order contribution to the dipole moment through $\bra{\Psi^{(1)}(t)}\mathbf{d}\ket{\Psi^{(1)}(t)}$), and the second term is the contribution from the second-order population of the excited states, which must be mediated by a virtual state (which in this model can only be another excited state).  We have kept both rotating and counter-rotating contributions.

We may further simplify Eq. \eqref{eq:static dipole toy model} by using a basis for the $N$ degenerate excited states in which the dipole operator is diagonal:
\beq
    \mathbf{d}_{ij} = \mathbf{d}_{ii}\delta_{ij}.
\eeq
In such a basis, we obtain

\begin{multline}
     \langle \mathbf{d}\rangle_t \overset{\text{static}}{=} \frac{1}{4}\left[ \left(\frac{1}{w_1-w_0 -\Omega}\right)^2 + \left(\frac{1}{w_1-w_0 +\Omega}\right)^2 \right] \sum_{i=1}^N |\mathbf{d}_{i0}\cdot\mathbf{E}|^2 \mathbf{d}_{ii} \\ + \frac{1}{w_1-w_0-\Omega}\frac{1}{w_1-w_0+\Omega} \sum_{i=1}^N \text{Re}\left[\mathbf{d}_{0i} (\mathbf{d}_{ii}\cdot\mathbf{E})(\mathbf{d}_{i0}\cdot\mathbf{E}) \right].
\end{multline}

\subsubsection{Floquet perturbation theory}

We present an alternate calculation of the static dipole moment using the Floquet formalism.  We consider the same model as in the previous section, now with the pulse width set to zero from the beginning, i.e.,
\beq
    H^{(1)}(t) = -\mathbf{d}\cdot \mathbf{E}\cos(\Omega t).
\eeq
The Floquet formalism directly accesses the regime of the broad middle of the wave train in time; in particular, the turning on and off of the pump pulse are not included in this calculation.  We consider the start time to be $t=0$ for convenience.

We now set up Floquet perturbation theory.  The Floquet Hamiltonian $H_F$ is an infinite-dimensional, Hermitian matrix defined by
\beq
    (H_F)_{ij,mn}= \frac{1}{\tau}\int_0^\tau dt\ e^{i(m-n)\Omega t} H_{ij}(t),
\eeq
where $\tau=1/\Omega$ is the period and where the Floquet indices $m$ and $n$ go over all integers.  With $H(t) = H^{(0)}+H^{(1)}(t)$, we obtain $H_F= H_F^{(0)}+H_F^{(1)}$ with

\bseq
\begin{align}
   (H_F^{(0)})_{ij,mn}&= (w_i - m \Omega) \delta_{ij}\delta_{mn},\\
   (H_F^{(1)})_{ij,mn}&= -\frac{1}{2} \mathbf{d}_{ij}\cdot\mathbf{E}(\delta_{m-n,1} + \delta_{m-n,-1}).
\end{align}
\eseq

By Floquet's theorem, there is a complete basis of time-evolving states of the form
\beq
    \ket{\Psi(t)} = e^{-i\epsilon t}\ket{u(t)},\label{eq:Floquet theorem}
\eeq
where $\epsilon$ is the quasienergy and where the time-evolving states $\ket{u(t)}$ are $\tau$-periodic.  We may write
\beq
    \ket{u(t)} = \sum_{i=0}^N\sum_m e^{-i m\Omega t}u_{im}\ket{i} = \sum_\alpha e^{-i \alpha_2\Omega t} u_\alpha \ket{\alpha_1},\label{eq:u(t)}
\eeq
where $\alpha = (i,m)$ is a two-component index consisting of the level index $i$ and the Floquet index $m$.  (Relative to the convention used in the main text, our convention here for the Floquet index differs by an overall sign.)

It is convenient to consider an auxiliary Hilbert space spanned by orthogonal basis vectors $|\alpha)$, where we write $|\alpha)$ instead of $\ket{\alpha}$ to avoid confusion with the original Hilbert space.  Then we write
\beq
    |u) = \sum_\alpha u_\alpha|\alpha).
\eeq
It may be shown that the quasienergies $\epsilon$ and the coefficients $u_\alpha$ are determined by an eigenvalue equation with the same structure as the time-independent Schrodinger equation:
\beq
    H_F |u) = \epsilon |u).
\eeq
We can thus apply standard time-independent perturbation theory to calculate $\epsilon = \epsilon^{(0)}+\epsilon^{(1)}+\dots$ and $|u)= |u^{(0)})+|u^{(1)})+\dots$, yielding the time-dependent wavefunction.

To determine the static dipole moment in the case that the ground state $\ket{0}$ is perturbed by light, we consider the case of $|u^{(0)}) = |0,0)$ and $\epsilon^{(0)}= w_0$.  The static dipole moment is given by

\bseq
\begin{align}
    \bra{\Psi(t)}\mathbf{d}\ket{\Psi(t)} &= \sum_{i,i'=0}^N \sum_m u_{(i',m)}^{*}\mathbf{d}_{i' i }u_{(i,m)}\\
    &= 2\sum_{i=1}^N\text{Re}\left[\mathbf{d}_{0 i }u_{(i,0)}^{(1)} \right] + \sum_{i,i'=1}^N \sum_m u_{(i',m)}^{(1)*}\mathbf{d}_{i' i }u_{(i,m)}^{(1)} + 2\sum_{i=1}^N \text{Re}\left[ \mathbf{d}_{0 i }u_{(i,0)}^{(2)}\right],\label{eq:static dipole in terms of u toy model Floquet}
\end{align}
\eseq
where, in the second line, we have expanded to second order and recalled $\mathbf{d}_{00}=\mathbf{0}$.

At first order, we readily obtain
\beq
    \epsilon^{(1)}= (0,0|H_F^{(1)}|0,0) = 0,
\eeq
and, for $\alpha \ne (0,0)$,
\beq
    u_\alpha^{(1)} = - \frac{(\alpha|H_F^{(1)}|0,0)}{\epsilon_\alpha - w_0}= \frac{1}{2}\mathbf{d}_{\alpha_1 0}\cdot\mathbf{E}
    \begin{cases}
        \frac{1}{w_1 - w_0 -\Omega} & \alpha_2 = 1,\\
        \frac{1}{w_1 - w_0 +\Omega} & \alpha_2 = -1,\\
        0 & \text{otherwise},
    \end{cases}
\eeq
where $\epsilon_\alpha = w_{\alpha_1}-\alpha_2\Omega$ is the unperturbed quasienergy of state $|\alpha)$.  The remaining coefficient, $u_{(0,0)}^{(1)}$, can be shown to be pure imaginary (due to the normalization requirement $\bra{\Psi(t)}\ket{\Psi(t)}=1$) and represents a phase degree of freedom in the choice of initial state; for convenience, we set $u_{(0,0)}^{(1)}=0$.  Note from Eq. \eqref{eq:static dipole in terms of u toy model Floquet} that there is no static dipole at first order, since $u_{i0}^{(1)}=0$ for $i=1,\dots,N$.

At second order, we obtain
\beq
    \epsilon^{(2)} = - \sum_{\alpha\ne(0,0)} \frac{|(\alpha|H_F^{(1)}|0,0)|^2}{\epsilon_\alpha-w_0}= -\frac{1}{4}\left(\frac{1}{w_1-w_0 -\Omega} +\frac{1}{w_1-w_0 +\Omega}\right)\sum_{i=1}^N |\mathbf{d}_{i0}\cdot\mathbf{E}|^2, 
\eeq
and, for $i=1,\dots,N$,
\beq
    u_{(i,0)}^{(2)} = \frac{1}{w_1-w_0}\sum_{\alpha\ne(0,0)} \frac{(i,0|H_F^{(1)}|\alpha)(\alpha|H_F^{(1)}|0,0)}{\epsilon_\alpha-w_0}= \frac{1}{2}\frac{1}{w_1-w_0-\Omega}\frac{1}{w_1-w_0+\Omega} \sum_{j=1}^N (\mathbf{d}_{ij}\cdot\mathbf{E})(\mathbf{d}_{j0}\cdot\mathbf{E}).
\eeq

Eq. \eqref{eq:static dipole in terms of u toy model Floquet} then recovers Eq. \eqref{eq:static dipole toy model}.  The second term on the right-hand side of Eq. \eqref{eq:static dipole in terms of u toy model Floquet} is the contribution of the first-order hybridization between $|0,0)$ and either $|i,1)$ or $|i,-1)$, while the third term is the contribution from the second-order hybridization between $|0,0)$ and $|i,0)$ mediated by either $|j,1)$ or $|j,-1)$ (both possibilities being summed, in addition to summing $i,j=1,\dots,N$). 

From Eqs. \eqref{eq:Floquet theorem} and \eqref{eq:u(t)}, we can also confirm that Floquet perturbation theory recovers the same time-dependent wavefunction as calculated in the previous section (in the limit of $t_0\to-\infty$ followed by $\Delta\to0^+$) except for the following two features.  First, the second-order correction to the quasienergy, $\epsilon^{(2)}$, did not appear in the previous section; however, in the second-order expansion of $\ket{\Psi(t)}$, $\epsilon^{(2)}$ can only appear as a (time-diverging) term in the coefficient $a_{0,I}^{(2)}$ that we did not need to calculate.  Second, there is an unimportant overall phase of $e^{-iw_0 t_0}$ between the two calculations, which arises from changing the switch-on time from $t_0$ to $0$.

If we wish to make the approximation of dropping the counter-rotating terms, we can do this directly by setting $\Omega\approx E_1-E_0$ in Eq. \eqref{eq:static dipole toy model}.  Alternatively, we can truncate the Floquet Hamiltonian from the beginning so that it acts on the basis states $|0,0), |i,1)$, and $|i,0)$ ($i=1,\dots, N$).  Then, the static dipole moment consists of a contribution from the first-order hybridization between $|0,0)$ and $|i,1)$, and a contribution from the second-order hybridization between $|0,0)$ and $|i,0)$ mediated by $|j,1)$.

\subsubsection{Equivalence to single-electron approach}
As an alternative to the calculations we presented above, that considered the multielectron states directly, we can instead work with single-electron states.  Note that, due to strong on-site Coulomb interaction, we can work with spinless electrons.  However, it seems that we must neglect the splitting between the $^{4}T_{1g}$ states and $^{4}T_{2g}$ states, because this is a more complicated interaction effect.

In the single-electron approach, we consider
\beq
    H^{(0)} = \sum_a w_a c_a^\dagger c_a, 
\eeq
where the index $a$ goes over five orthonormal, single-electron states: the three $t_{2g}$ states (which we label as $g_1,g_2,g_3$) and the two $e_g$ states ($e_1,e_2$).  We are interested in the time-evolving dipole moment starting from the following three-electron state:
\beq
    \ket{\Psi} = c_{g_3}^\dagger c_{g_2}^\dagger c_{g_1}^\dagger\ket{\Omega},
\eeq
where $\ket{\Omega}$ is the empty state (annihlated by all $c_a$).  The state $\ket{\Psi}$ is the same as the ground state $\ket{0}=\ket{g}$ considered earlier.

The coupling to the pump field is as before.  Now we note that the dipole operator is quadratic in the single-electron fields:
\beq
    \mathbf{d} = \sum_{a,a'} c_{a'}^\dagger\mathbf{d}_{a' a}c_a^\dagger.
\eeq

It is then straightforward to show that the time-evolving dipole moment is the sum of time-evolving dipole moments in the three single-electron states in $\ket{\Psi}$.  That is,
\beq
    \bra{\Psi}U^\dagger(t,t_0) \mathbf{d} U(t,t_0)\ket{\Psi} = \sum_{a_0= g_1,g_2,g_3}  \bra{\Omega} c_{a_0} U^\dagger(t,t_0) \mathbf{d} U(t,t_0)c_{a_0}^\dagger \ket{\Omega}.\label{eq:dipole as single-electron} 
\eeq
We can show Eq. \eqref{eq:dipole as single-electron} by, e.g., using the time-dependent creation operators.  In particular, we may define
\beq
    c_a^\dagger(t) = U^\dagger(t,t_0)c_a^\dagger U(t,t_0),
\eeq
where we note that the time evolution is in the opposite sense compared to the Heisenberg picture.  We then have
\beq
    U(t,t_0)\ket{\Psi} = c_{g_3}^\dagger(t)c_{g_2}^\dagger(t)c_{g_1}^\dagger(t) \ket{\Omega}.
\eeq
Noting that the canonical anticommutation relation holds at all time ($\{ c_a(t),c_{a'}^\dagger(t)\} = \delta_{aa'}$), and noting that the dipole operator is quadratic, we then obtain
\beq
    \bra{\Psi}U^\dagger(t,t_0) \mathbf{d} U(t,t_0)\ket{\Psi} = \sum_{a_0=g_1,g_2,g_3} \sum_{a,a'}\{ c_{a_0}(t), c_{a'}^\dagger\}\mathbf{d}_{a' a} \{c_a, c_{a_0}^\dagger(t) \},
\eeq
which is equivalent to Eq. \eqref{eq:dipole as single-electron}.  

Thus, an alternate approach to calculating the dipole moment is to add up the dipole moments in the three single-electron problems corresponding to the three $t_{2g}$ states.

\section{Discussion on other mechanisms of the symmetry breaking }
We have demonstrated that the observed ultrafast symmetry breaking in  Cr$_2$O$_3$ with linearly polarized light originates from a light-induced electronic dipole moment via optical rectification. Such an effect has purely electronic origin as can be seen from the ultrafast timescale of the response. The interpretation of the data in terms of a light-induced electronic dipole moment is also supported by several experimental features, such as the opposite response to the pump in the two AFM domain states, the pump polarization dependence, the sensitive wavelength and temperature dependence, as well as the absence of a response to a circularly polarized pump. Despite the excellent agreement between the experiment and the interpretation presented, in this section, we further rule out other possibilities that may potentially contribute to the effect.

\subsubsection{Sum frequency generation (SFG) and difference frequency generation (DFG)}
In this section, we specifically address the possibility that SFG/DFG between the pump and the probe could yield the signatures observed. We present below three arguments that show that this is not a viable interpretation.

First, coherent effects such as SFG/DFG could deplete the probe beam in a polarization-dependent fashion. A fraction of the photons from the probe beam would then no longer be available to generate SHG. It is important to note that such a process can only decrease the probe beam intensity and can thereby only decrease the SHG signal. In contrast, at certain polarization angles, we observe a large increase of the SHG signal, up to $\sim$30\%, which is a strong evidence against the SFG/DFG interpretation. 

Second, the calculated SFG/DFG polarization dependence is inconsistent with the observed light-induced symmetry-breaking. The polarization dependence of the SFG signal (which is the same for DFG) can be calculated using $\boldsymbol{P}_{SFG}(2\omega)=\boldsymbol{\chi} \boldsymbol{E}_{pump} (\omega)\boldsymbol{E}_{probe} (\omega)$ with $\boldsymbol{E}_{probe}=E_{probe} \textrm{sin}(\theta) \textbf{e}_{x} +  E_{probe} \textrm{cos}(\theta) \textbf{e}_{y}$, $\boldsymbol{E}_{pump}=E_{pump} \textrm{sin}(\phi) \textbf{e}_{x} +  E_{probe} \textrm{cos}(\phi) \textbf{e}_{y}$, where $\boldsymbol{\chi}$ is the second order susceptibility tensor constrained by the $\underline{\bar{3}m}$ symmetry of Cr$_2$O$_3$. We get that $ \boldsymbol{P}_{SFG}(2\omega) \propto E_{pump} E_{probe} (-\textrm{sin}(\theta+\phi) \textbf{e}_{x} +\textrm{cos}(\theta+\phi) \textbf{e}_{y}) $. This expression implies that the intensity of the SFG beam, $I_{SFG} \propto |\boldsymbol{P}_{SHG}|^2$ is independent of both the pump and probe polarization; the generated SFG is isotropic. Our calculation is consistent with our experimental observations, where we observe an overall isotropic intensity decrease at time zero. An isotropic scaling factor of 0.9 was required to multiply the equilibrium susceptibility parameters when fitting the pumped SHG patterns. We ascribe this effect to the SFG/DFG process.

Third, The fluence dependence of our data (Fig.4(b) in the main manuscript) also corroborates our interpretation. The symmetry-breaking effect, which can be quantified by $\delta\chi^e$ in the manuscript, is clearly proportional to the pump fluence (i.e. proportional to $|E|^2$). However, the effect would be proportional to the pump electric field if the observed effect originated from a two-photo SFG/DFG process. Hence, depletion of the probe beam due to SFG/DFG is inconsistent with our fluence dependence. For three photon SFG (one photon from the probe and two photons from the pump), $\delta\chi^e$ may be proportional to fluence. However, this three-photo process is also allowed above $T_N$, and we do not observe any symmetry reduction above $T_N$ as shown in Fig.~S1. We thus deem this effect negligible.

\subsubsection{Incoherent decay (luminescence)}
Because the pump light energy shone is less than the energy of the lowest lying electronic excited state, there is almost no absorption of the pump light and therefore almost no incoherent decay. The sample is largely transparent to the pump light. Additionally, an incoherent decay/luminescence would normally lead to an asymmetric lineshape in the time trace, while we observe a largely symmetric lineshape with a FWHM of $\sim$180fs. 

The slight asymmetry in the background level is due to an incoherent response – two-photon absorption. However, the symmetry-breaking response of the solid is only present while the pump pulse is present (Fig. 2(a) and (c)), and therefore the small incoherent response cannot be responsible for the observed symmetry-breaking effect.

\subsubsection{Three-photon processes}
Three photon processes are generally not constrained by inversion symmetry of the solid, and it is allowed above T$_N$ as well. In our experiments, the symmetry-breaking response is only observed below T$_N$, when inversion symmetry is broken (Fig.~S1), therefore the three-photo processes are negligible in our experiment.

\subsubsection{Heating due to two-photon excitations}
By pumping below the gap, direct excitations are avoided, hence light-induced photo-currents and heating are significantly minimized. Luminescence, hot carriers and hot plasmonic states are thus negligible. In addition, the timescale of these responses are typically longer than 200~fs and exhibit asymmetric time evolution. Therefore, we expect that such effects are not relevant to the observations presented here. 

The dominant source of heating in our case is indirect excitations such as two-photon excitations. Multi-photon excitations are inevitable in all experiments with intense lasers. As shown in Fig~2(a) and (c), our time traces exhibit  a slight asymmetry in the background level ($\sim$500 fs before and after time zero), which is due to two-photon excitations. If there were no such multiphoton excitations, our time trace would be perfectly symmetric. Nevertheless, the symmetry-breaking response is symmetric about zero time delay, and therefore the asymmetry in the background is unrelated to the symmetry breaking at zero time delay. Instead, the multi-photon absorption gives rise to an isotropic reduction of the SHG signal, which can be extracted due to the much longer relaxation timescale.

If this mechanism, transient photocurrents, or a superheated transient state were responsible for our observations, they should have resulted in a longer timescale compared to those of our observations. In both cases, one would have expected a relaxation timescale on the order of several hundred femtoseconds to picoseconds. Furthermore, one would have expected an enhanced temporal asymmetry in the symmetry-breaking response. Therefore, none of these effects can be responsible for our observations.

The heating in our experiments is greatly reduced by pumping below the gap, and in fact the intense pump laser only increases the temperature by roughly 5~K. The SHG intensity is different when probing with left and right circularly polarized light due to the interference between the electric and magnetic dipole SHG radiation. Since the electric dipole SHG is absent above T$_N\sim$307~K, the distinction between left and right circular light disappears above $T_{N}$. We can utilize this to estimate the temperature increase due to the pump. When we pump the sample at 300~K, there is still distinction between left and right circular light, indicating the temperature is still below $T_{N}$ in the presence of the pump. The temperature dependence of the pumped anisotropic SHG patterns also corroborate the small temperature increase by the pump.

\section{How symmetry-breaking in $\textrm{Cr}_2\textrm{O}_3$ is different from that in equilibrium electronic nematics}

There are several fundamental ways in which the light-induced electronic symmetry-breaking in Cr$_2$O$_3$ differs from that in conventional electronic nematics. First,  in conventional electronic nematics, in-plane rotational symmetry enforces a degeneracy of the ground state that is broken upon cooling below the nematic transition temperature. In our work on Cr$_2$O$_3$, it is not the degeneracy of the ground state that is relevant but how the degenerate excited states are unequally superposed with the ground state when dressed with light. The polarization of the pump light selects out only part of the degenerate excited state manifold to mix in with the ground state wavefunction to generate a non-zero dipole moment. For example, because the excited state manifold here comprises states that are equal superpositions of Cr $3d_{xy}$  and $3d_{x^2-y^2}$ orbitals, $3d_{xz}$ and $3d_{yz}$ orbitals, as well as Cr $4p_x$ and $4p_y$ orbitals (see Supplemental Note VIII), the electron probability density is three-fold symmetric in the plane, as required by the crystal symmetry. When intense light propagates through Cr$_2$O$_3$, the polarization of light selects out part of this excited state manifold to superpose with the ground state, which breaks the in-plane threefold rotational symmetry. Second, the timescale observed for the light-induced symmetry-breaking transition indicates that the electronic degree of freedom decouples from the lattice. In conventional electronic nematics, isolating the electronic from the lattice subsystem is usually not possible. Last, because the symmetry reduction here is related to the polarization of the pump light, the orientation of the broken symmetry is controllable with an external perturbation in contrast to conventional electronic nematics that are pinned by the crystal lattice.

\section{Comparison with the work by $\textrm{Sala et al.}$ (Ref. 33 of main text)}
A pump-probe experiment on Cr$_2$O$_3$ using SHG was previously published by Sala et al., which showed qualitatively similar results, including similar RA-SHG patterns. In their work, the authors assigned the observed symmetry reduction to a lattice distortion and demagnetization process. In this section, we draw comparisons between our work and that of Sala et al., which consist of experimental and interpretational distinctions. 

One of the central points of our work is that when the pump light possesses an energy below the electronic gap, the electronic symmetry is lowered while the spins and lattice remain largely unperturbed -- we are able to isolate the electronic response. In our experiment, the symmetry reduction is therefore approximately symmetric over the width of the laser pulse ($\sim$180 fs). This observation demonstrates that the response is quasi-instantaneous within our temporal resolution and indicative of a purely electronic response. However, in the work of Sala et al, the laser pulse widths used were 35~fs, and the symmetry-breaking response was observed for approximately 400~fs (this is observed when the pump is tuned to 1.8~eV, i.e. resonantly pumping a $d-d$ excitation). The symmetry-breaking response was not observed only over the duration of the pulse width, but also long after the pulse was gone.

Because of the different response timescales observed in the two works, the interpretations also differ; Sala et al. attributed their observations to a lattice response. In addition to the timescale observed by Sala et al., the polarization dependence of the RA-SHG patterns was also interpreted as a lattice distortion of either $A_{1g}$ or $E_g$ symmetry. Because photons were absorbed in their experiment (the lowest energy pump light used was 1.8~eV), it is difficult to disentangle the electronic from the structural response, and their assignment of an $A_{1g}$ or $E_g$ distortion may well be correct for the structural component. An important implication of the assigned symmetry of the structural response is that it cannot discriminate between different AFM domain states. The response of the opposite domain state was not investigated in the work by Sala et al. Our higher angular resolution data and our data in opposite AFM domain states allow us to draw the conclusion that, in our work, the transient symmetry breaking is due to a light-induced electronic dipole moment. This interpretation is supported by many experimental features in our work, including the different responses in the two AFM domain states (Fig.~2(b) and (d)), the pump polarization dependence (Fig.~3(a)), the temperature dependence (Figs.~S1 and S2), the wavelength dependence (Fig.~S6), and the fluence dependence (Fig.~4(b)). Importantly, our perturbation is odd under parity, while the structural response symmetry determined by Sala et al. is even (either $A_{1g}$ or $E_g$).
 
An electronic response was also likely present in the work of Sala et al., though the authors did not present evidence or come to that conclusion in their publication. Even if an electronic response was present, the driving mechanisms can be different when comparing a sub-gap to a resonant drive. When resonantly pumping an electronic excitation, absorption is strong and energy is transferred from the light field to the solid. In this regime, optical rectification must be divided into three distinct mechanisms – direct optical rectification, a shift current and an injection current. All three mechanisms are “electronic”, but they are not the same. If an electronic mechanism is present in the work of Sala et al, it must be through either a shift or injection current (most likely a shift current), which can potentially last longer than the width of the laser pulse. In our work, only direct optical rectification is possible, because the pump is sub-gap with minimal absorption. Resonantly pumping versus pumping below the gap leads to a qualitative distinction – the physical mechanisms are different. In our study, we work in a regime where only direct optical rectification is possible and the electronic response can be isolated.